\documentclass[manuscript]{aastex}

\usepackage{graphicx}
\usepackage{rotating}
\usepackage{dcolumn}

\newcommand{\rsun}{R_{\Sun}}
\newcommand{\mydeg}{^{\circ}}

\shorttitle{CME Deflection}
\shortauthors{Kay et al.}

\begin{document}

\title{Global Trends of CME Deflections Based on CME and Solar Parameters}

\author{C. Kay}
\affil{Astronomy Department, Boston University, Boston, MA 02215}
\email{ckay@bu.edu}

\author{M. Opher}
\affil{Astronomy Department, Boston University, Boston, MA 02215}

\and 

\author{R. M. Evans}
\affil{NASA Goddard Space Flight Center, Space Weather Lab, Greenbelt, MD, 20771}

\begin{abstract}
Accurate space weather forecasting requires knowledge of the trajectory of coronal mass ejections (CMEs), including any deflections close to the Sun or through interplanetary space. \citet{Kay13} introduced ForeCAT, a model of CME deflection resulting from the background solar magnetic field. For a magnetic field solution corresponding to Carrington Rotation (CR) 2029 (declining phase, April-May 2005), the majority of the CMEs deflected to the Heliospheric Current Sheet (HCS), the minimum in magnetic pressure on global scales.  Most of the deflection occurred below 4 $\rsun$.  Here we extend ForeCAT to include a three dimensional description of the deflecting CME.  We attempt to answer the following questions: a) Do all CMEs deflect to the magnetic minimum? and b) Does most deflection occur within the first few solar radii (~4 $\rsun$)?  Results for solar minimum and declining phase CMEs show that not every CME deflects to the magnetic minimum and that the deflection is typically determined below 2 $\rsun$.  Slow, wide, low mass CMEs in declining phase solar backgrounds with strong magnetic field and magnetic gradients exhibit the largest deflections.  Local gradients related to active regions tend to cause the largest deviations from the deflection predicted by global magnetic gradients, but variations can also be seen for CMEs in the quiet sun regions of the declining phase CR.  We show the torques due to differential forces along the CME can cause rotation about the CME's toroidal axis. 
\end{abstract}


\section{Introduction}
Plasma explosions known as coronal mass ejections (CMEs) routinely erupt from the Sun's surface.  CMEs drive space weather phenomena at Earth and throughout the rest of the heliosphere.  Energetic particles accelerated by CME-driven shocks can harm astronauts at Earth and damage satellites throughout the solar system.  Forecasting space weather effects relies on knowledge of the path of a CME.  Observations commonly show significant non-radial deviations in the CME trajectories.  Understanding these deflections will allow for more accurate space weather predictions.

Measurements of CME deflections come from coronagraph observations.  Initially only latitudinal CME deflections were observed within a coronagraph image from a single viewpoint \citep{Mac86, Cre04, Kil09}.  Recently longitudinal deflections have been measured by reconstructing the three dimensional CME trajectory using coronagraph or heliospheric imager observations from multiple viewpoints \citep{Byr10,Liu10a,Liu10b,Lug10,Isa13, NC14, Wan14}.  Coronal observations show that CMEs can undergo significant deflection close to the Sun, but it is often hard to disentangle the effects of deflection, rotation, and nonuniform expansion in the lower corona \citep{NC12}.  \citet{Byr10} measure a latitudinal deflection of 30$\mydeg$ below 7 $\rsun$ for the 2008 December 12 CME.  \citet{Isa13} reconstruct 15 CME trajectories and find that the magnitude of latitudinal deflections tends to exceed the magnitude of longitudinal deflections.  The 15 CMEs reconstructed by \citet{Isa13} show an average longitudinal deflection of only 1.7$\mydeg$ compared to an average latitudinal deflection of 14.3$\mydeg$.

Many mechanisms have been proposed to explain coronal CME deflection (see \citet{Kay13} and references within).  The present work focuses on deflection due to magnetic forces.  \citet{Kil09} suggest that CMEs may not be able to penetrate the open magnetic field emanating from coronal holes (CHs).  The CH magnetic field then guides CMEs toward the Heliospheric Current Sheet (HCS).  \citet{She11} and \citet{Gui11} attribute the deflection to gradients in the background magnetic energy density, which would also cause CMEs to tend to deflect toward the HCS.  At solar minimum, the HCS remains flat at low latitudes, so predominantly latitudinal deflections should occur as a result of magnetic forces.  During other times of the solar cycle, the HCS transitions to a more complex configuration allowing deflections to have a more significant longitudinal component.  The 15 CMEs of \citet{Isa13} occurred during solar minimum (between 2008 and 2010) so magnetic forces would explain the observed smaller longitudinal component of the deflections.  

CME deflections also occur in magnetohydrodynamic (MHD) simulations \citep{Lug11, Lug12,Zuc12, Lyn13, Zho13, Zho14}.  As with the observed CMEs, the MHD CMEs tend towards regions of lower magnetic energy.  In some cases magnetic reconnection creates an imbalance in the magnetic energy which causes a CME to deflect early in the eruption \citep{Zuc12, Lyn13}.  MHD simulations also show that CMEs can deflect due to interactions with other CMEs \citep{Lug12}.

In \citet{Kay13}, hereafter referred to as K13, we introduced ForeCAT (Forecasting a CME's Altered Trajectory), a versatile model for CME deflection.  Given a magnetogram and the initial CME characteristics, ForeCAT uses a simplified solar background to predict the expected deflection of the CME.  ForeCAT includes the radial propagation and expansion of the CME as well as the background magnetic forces that deflect the CME.  Because ForeCAT is optimized to be computationally efficient, as opposed to an MHD model, we are easily able to run a large number of cases.  

In K13, we explored CMEs erupting from the active region (AR) 0758 in Carrington Rotation (CR) 2029.  This background contains strong gradients in the magnetic field strength as the AR is located in close proximity to both a low latitude CH and the streamer belt (SB) that transitions into the HCS at farther distances.  The strong global magnetic gradients in this background caused the majority of the simulated CMEs to deflect to the HCS, the minimum in the magnetic field strength, within the first few solar radii.  We describe this CME motion as deflection toward the ``magnetic minimum.''  

Close to the Sun the magnetic minimum tends to correspond to local null points resulting from the orientation of the SB and any AR within the streamer region.  At farther distances, the magnetic minimum corresponds to the HCS.  In the strong background explored in K13, most CMEs deflected to the SB magnetic minimum within the first few solar radii. The CMEs remained at the SB minimum as it transitioned to the HCS minimum and exhibited negligible deflection as the CMEs continued propagating to 1 AU.  If the SB and HCS minimum do not correspond to the same position, such as would be the case for a CME erupting within a pseudostreamer, we expect additional deflection to the HCS minimum at further distances.  Additionally, if the local AR magnetic gradients initially deflect the CME, a change in the direction of deflection may occur when the global gradients begin to dominate the local gradients.  We define global gradients as those resulting from the relative orientation of CHs and the HCS.  K13 explored deflection due the effects of the global background rather than local effects.  Here we consider the effect of both local and global magnetic gradients.  However our simplified background with a static magnetic field more accurately describes the global gradients since the local gradients evolve on shorter time scales.

Works such as \citet{Pan11} and \citet{Pan13} show that local gradients at ARs might be important.  For example, filaments can undergo significant deflections or rolling motions before the formation of the CME flux rope that encapsulates the filament.  Reproducing the rolling motion of filaments requires accurately describing the temporal evolution of local gradients that can evolve on time scales similar to the motion of the filament (of order hours).  The global gradients also evolve in time, but this change occurs much slower than the time necessary for a CME to propagate through the corona.  Future work with ForeCAT will address the temporal evolution of the local magnetic gradients, including those of ARs, which can cause variation in the initial non-radial motion.

Here, we present results from an expanded version of ForeCAT that follows the full three-dimensional motion of a deflecting CME.  We extend the analysis of K13 to multiple CRs to look at variations in the deflections of CMEs throughout the solar cycle.  Based on the conclusions of K13 we ask if all CMEs deflect to the magnetic minimum and if this deflection always occurs below 4 $\rsun$.  Section \ref{ForeCAT} describes the new version of ForeCAT and section \ref{sw} describes the solar backgrounds used in this work.  In section \ref{FCdefs} we present results for different CRs, and section \ref{modelsens} shows the sensitivity of these deflections to the radial propagation, non-radial drag, and magnetic field models.  Section \ref{CMEinputs} shows the effect of the initial CME parameters on the ForeCAT deflections and section \ref{CRmaps} shows the range of deflections for each CR.  Finally, in section \ref{DisCon}, we present conclusions and discussion.

\section{ForeCAT}\label{ForeCAT}
K13 introduced ForeCAT, which previously determined the deflection of a cross section of a CME within a two-dimensional ``deflection plane.''  The primary difference from the K13 version of ForeCAT is that we now determine the full three-dimensional motion.  The CME is no longer restricted to a single deflection plane.  We also utilize a two-dimensional grid on the surface of the three-dimensional CME and have made changes to the CME expansion model and the background solar wind and magnetic field model.  Additionally, we can include CME rotation resulting from differential forces along the CME axis.  Comparison with the results of K13 and the effect of each change can be found in appendix \ref{FC1comp}.

Figure \ref{fig:schem} contains a diagram representing the organization of ForeCAT.  ForeCAT requires descriptions of both the CME and the background solar wind, as shown on the left hand side of the diagram.  From these, ForeCAT calculates a CME's deflection using the background solar magnetic forces.  ForeCAT also includes the radial propagation of the CME, the expansion of the CME, and any drag affecting the CME's motion.  These four components allow for determination of the CME trajectory between the surface of the Sun and 1 AU.  We briefly describe the individual components in the following subsections.

\begin{figure}
\includegraphics[scale=0.25]{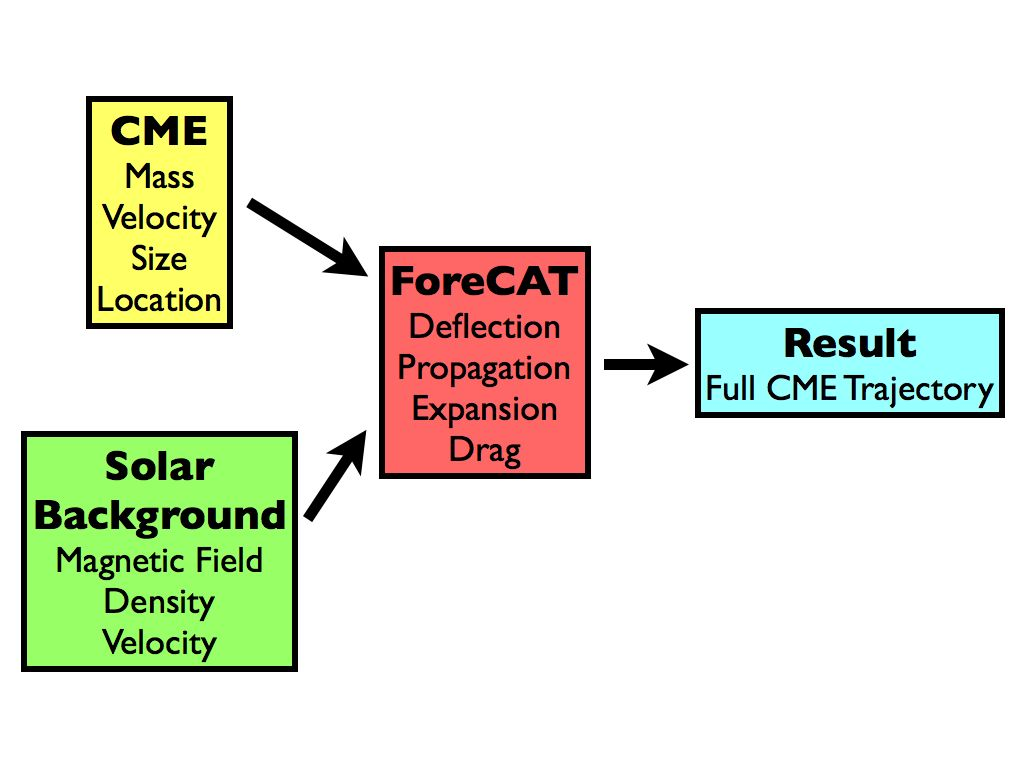}
\caption{Diagram showing the organization of ForeCAT. ForeCAT requires both CME parameters and a description of the solar magnetic field and solar wind.  In addition to determining the deflection, ForeCAT utilizes models for the CME's radial propagation and expansion, and determines the drag from the background solar wind.  The combination of these models produces the three-dimensional trajectory of a CME between 1 $\rsun$ and 1 AU.}\label{fig:schem}
\end{figure}

\subsection{CME Description}\label{CME}
ForeCAT represents the flux rope of a CME using a 3D torus that we assume persists out to 1 AU.  This representation of a flux rope is similar to many other models, including those of \citet{Gib98, TD99, Che96} and \citet{The06}.  ForeCAT makes no explicit assumptions about the CME magnetic field.  Observations of magnetic clouds (MCs) with smoothly rotating magnetic field suggest that CMEs can maintain a flux rope like structure out to 1 AU \citep{Bur81, KB82, CR03}.  \citet{Van02} perform an MHD simulation of a flux rope between 30 $\rsun$ and 1 AU.  Their flux rope shows some deformation but retains a predominantly torus-like structure out to 1 AU and their simulated in situ observations reproduce the characteristics of a MC.  Additionally, observations of the flux rope from a single CME at multiple spacecraft \citep{Mos12} support the presence of an extended torus throughout a CME's propagation.

\begin{figure}
\includegraphics[scale=0.13]{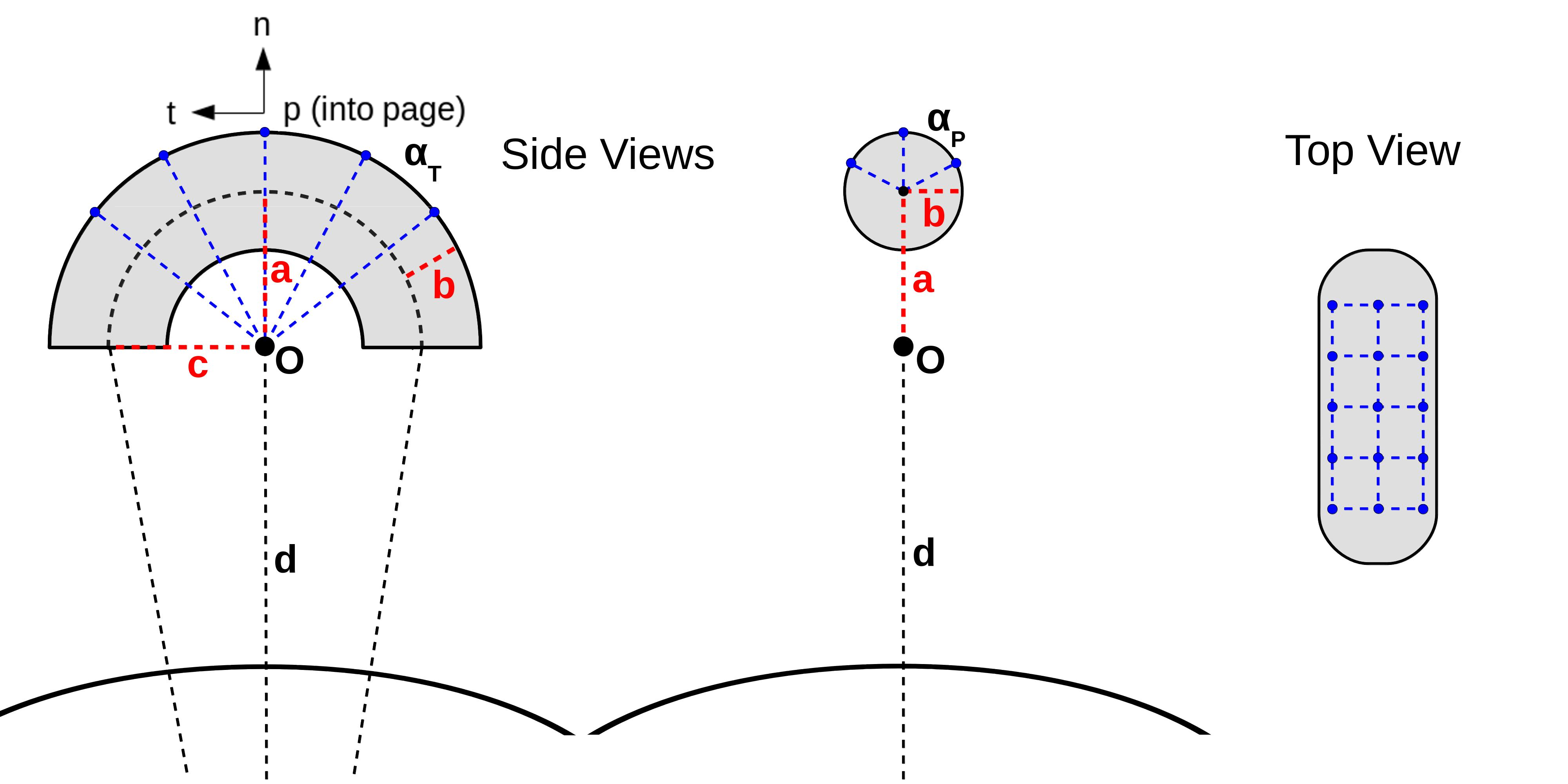}
\caption{Diagram showing two side views and one top view of ForeCAT's toroidal flux rope structure (gray shaded region). The torus-shape is defined by the height, $a$, width, $c$, and cross-sectional width $b$.  The parameter $d$ describes the distance between the center of the Sun and the origin, $O$, of the torus.  The blue dots indicated the grid points at which ForeCAT calculates the deflection forces.  The angles $\alpha_T$ and $\alpha_P$ determine the spacing of the grid points in the toroidal and poloidal directions and the coordinate axes show the toroidal (t), poloidal (p) and normal (n) directions. }\label{fig:diag}
\end{figure}

Figure \ref{fig:diag} shows ForeCAT's toroidal flux rope structure.   The toroidal axis traces out a half-ellipse (left panel) and the torus has a circular cross section.  The initial CME shape is completely specified by three parameters: $a$, $b$, and $c$.  The parameters $a$ and $c$ represent the axes of the ellipse formed by the toroidal axis- $a$ is defined as the axis in the direction of the nose of the CME (the point on the surface of the torus with greatest radial distance) and $c$ is in the perpendicular direction (toward the flanks).  The cross section is described by the radius $b$.  The parameter $d$ specifies the initial radial distance from the center of the Sun to the center of the CME.  

ForeCAT represents the flux rope structure numerically as a grid on the face of the torus with fifteen points in the toroidal direction and thirteen points in the poloidal direction for a total grid consisting of 195 points (blue dots in Fig. \ref{fig:diag}).  The deflection is sensitive to the actual number of points, however the solution converges for grids as large as 15x13, which is the grid we use throughout this work.  

The poloidal direction corresponds to the angular polar coordinate of the circular cross section.  The toroidal direction parallels the toroidal axis that intersects the center of every circular cross section (the origin of the polar coordinate system).  The normal direction is perpendicular to the torus surface.  This coordinate system is shown in the left panel of Fig. \ref{fig:diag}.  The center grid point is at the nose of the CME so that the grid is symmetric about the nose of the torus.  We do not include any points on the side of the CME closest to the Sun as our steady state solar wind description will not accurately describe the region behind the CME that has been perturbed from the steady state by the passage of the CME.

The initial torus position is specified by the latitude and longitude of the nose and a tilt angle that defines the orientation.  We define the tilt angle as the angle between the plane containing the toroidal CME axis and the equatorial plane so that a tilt of 0$\mydeg$ corresponds to a horizontal flux rope and a tilt of 90$\mydeg$ corresponds to a vertical flux rope.  The angle is defined so that positive angles correspond to counterclockwise rotation away from a horizontal flux rope.  When initiating CMEs in ForeCAT we determine the initial position, orientation, and length ($c$) by aligning the initial torus with polarity inversion lines (PILs), sites where flux ropes would likely form.  Without detailed observations of the initial flux ropes we approximate $a$ as equal to $c$ and set $b$ to 0.2 $c$ and $d$ to 1.05 $\rsun$.  We explore ForeCAT's sensitivity to the initial shape in section \ref{shape_vars}.

\subsection{CME Propagation and Expansion and Drag}\label{PED}

ForeCAT uses a three-phase radial propagation model to describe the CME's radial motion, similar to that of \citet{Zha06}.  The CME begins with a slow rise phase at constant velocity, $v_g$, followed by rapid acceleration, and finally constant radial propagation at velocity, $v_{CME}$.  The transitions between the phases occur at fixed radial distances, respectively $r_{ga}$ and $r_{ap}$.  These parameters can be adjusted to reproduce the measured radial velocity of individual observed cases.  Here we set $v_g$ to 80 km s$^{-1}$ and $r_{ga}$ to 1.5 $\rsun$ and $r_{ap}$ to 3 $\rsun$ and we explore ForeCAT's sensitivity to these values in section \ref{propsens}.

We do not explicitly include any drag effects in the radial direction.  The empirical form of the three-phase propagation model is determined from coronagraph observations of the motion of CMEs that implicitly incorporate the effects of radial drag.

The CME description is inspired by the ``ice-cream cone'' model of CMEs, which is frequently used to describe the shape and evolution of CMEs in white-light coronagraph images \citep{Fis84, Xie04, Xue05, Wan06}.  These models fit the bright, dense piled-up solar wind in front of the dark, tenuous flux rope.  Within coronagraph images, the CMEs generally appear to move with a fixed angular width and can be well-described by a cone.  This model describes a self-similar expansion where the CME expands proportionally with height producing a constant angular width, which observations commonly show occurring above about 5 $\rsun$ \citep{Che96, Che97, Woo09, Mie11}.  \citet{Low82, Low84} and \citet{Che96} show that self-similar expansion is an analytic solution to an MHD description of CME dynamics.   

Observations of CME expansion in the low corona suggest that CMEs may form from small-scale features that rapidly overexpand (faster than self-similar) into ``CME-scale'' features \citep{Che00, Cre04, Pat10a, Pat10b}.  However, this phase of rapid overexpansion may conclude at distances much closer than 5 $\rsun$ as self-similar expansion is also observed in the low corona \citep{The06, Woo09}, occasionally as close as 1.1 $\rsun$ \citep{Asc09}.  

Here we assume ForeCAT CMEs expand self-similarly.  The CME shape parameters, $a$, $b$, and $c$ scale linearly with the CME distance, $d$.  The previous version of ForeCAT used the Melon Seed OverExpansion (MSOE) model \citep{Sis06}.  The MSOE model produces overexpansion in the low corona, however, the resulting expansion tends to exceed that of observations (A. Vourlidas, personal communication).  ForeCAT is highly flexible so that the expansion and propagation models can easily be replaced with either empirical relations or more sophisticated physics-based models.  In the future, ForeCAT will use physics-based models to determine the propagation and expansion from the magnetic field of the CME but to begin understanding the general properties of CME deflections we proceed with these simplified models.

ForeCAT includes the effects of drag only in the non-radial direction.  The three-phase propagation model describes the CME's radial dynamics so we do not include any drag in the radial direction.  ForeCAT determines the drag force per unit length, $\frac{f_D}{\ell}$, following the form of \citet{Car96} and \citet{Car04},  
\begin{equation}
\frac{f_D}{\ell} = -C_d b \rho_{SW} (\vec{v}_{CME,nr} - \vec{v}_{SW,nr})| \vec{v}_{CME,nr} - \vec{v}_{SW,nr}|
\end{equation}
where $C_d$ is the drag coefficient, $b$ is the cross-sectional radius of the CME, and $v_{CME,nr}$ and $v_{SW,nr}$ are respectively the non-radial velocities of the CME and solar wind.  Given the CME volume, $\pi b^2 \ell$, we determine the volumetric force.
\begin{equation} \label{drageq}
F_D = -\frac{2 C_d \tanh \beta \rho_{SW}}{\pi b} (\vec{v}_{CME,nr} - \vec{v}_{SW,nr})| \vec{v}_{CME,nr} - \vec{v}_{SW,nr}| 
\end{equation}
The value of the drag coefficient is unknown but values near unity are typically used.  \citet{For06} show that a constant drag coefficient results in CME velocities in the low corona that do not match observations but the observations can be reproduced by setting with a drag coefficient having the profile of $\tanh \beta $, where $\beta$ is the plasma beta.  We set the drag coefficient to be equal to $C_d \tanh \beta$ where $C_d$ is a constant near unity.  We the following form for $\beta$ versus distance determined from Fig 1.22 of \citet{AscB}
\begin{equation}
\beta (R) = 2.515 (R -1)^{1.382}
\end{equation} 
where R has units of solar radii.  This expression is determined from coronal values and rapidly increases for large values of $R$.  However, since we take the hyperbolic tangent of $\beta$ the drag coefficient is equivalent to one above 10 $\rsun$ so it does not effect ForeCAT results. The actual value of $\beta$ and its dependence on distance should vary between ARs, CHs, and the quiet sun.  For now ForeCAT uses the above expression to limit the number of free parameters.  Future work will incorporate thermal pressure gradients into ForeCAT and $\beta$ will be calculated as the ratio of the thermal and magnetic pressure when ForeCAT includes a full thermodynamic description of the solar background.

\subsection{Deflection Forces}
The magnetic deflection force is determined from both components of the Lorentz force: magnetic pressure gradients and magnetic tension.  We use a static model of the solar magnetic field and approximate the draping of the magnetic field around the CME to determine the Lorentz force from the background magnetic field.  We do not include any enhancements in the magnetic field from the CME expansion compressing the surrounding solar wind, making our forces lower limits, nor any rotation in the magnetic field due to CME driven shocks \citep{Liu11}.  The magnetic forces which deflect the CME become negligible by 2 $\rsun$ where it is not clear that shocks will form at these low heights. 

\subsection{CME Trajectory}\label{traj}
The deflection forces, $\vec{F}$, are volumetric forces that describe the acceleration, $\vec{a}$, of a single point within a fluid with density $\rho$. 
\begin{equation}\label{Fpa}
\vec{F} = \rho \vec{a}
\end{equation}
This equation must be integrated over the full CME volume to determine the acceleration of the CME mass, $M_{CME}$.  With discretized grid points, the integral over the CME volume can be expressed as a sum of the volumetric forces multiplied by the corresponding volume element, $\Delta V$.
\begin{equation}
\int \vec{F} dV \approx \sum \vec{F} \Delta V = M _{CME} \; \vec{a}
\end{equation}
This requires summing the forces over the full volume of the CME.  ForeCAT only determines the forces on the front surface of the CME so instead, we determine the average volumetric force from the surface forces.  We then determine the acceleration using Eq. \ref{Fpa} with the average CME density.  Our drag force does not vary with location so that the average volumetric drag force is described by Eq. \ref{drageq} and the non-radial CME acceleration is
\begin{equation}\label{acceleq}
\vec{a} = \frac{1}{\rho_{CME}} \left(\frac{1}{N}\sum\limits^N{(\vec{F}_{G} + \vec{F}_{T})} + \vec{F}_D\right)
\end{equation}
where $N$ is the number of grid points, in this case 15. Eq. \ref{acceleq} combined with the radial propagation model, determines the motion of the CME.  The new position of the CME nose is calculated as
\begin{equation}
\vec{x}(t+ \Delta t) = \vec{x}(t) + (\vec{v}_{nr}(t) + \vec{v}_{prop}(t)) \Delta t + 0.5 \vec{a}(t) \Delta t^2
\end{equation}
where $t$ and $t+ \Delta t$ refer to the current and following time step and $\vec{v}_{nr}$ is the non-radial velocity representing the combined effects of deflection and drag.  We use a time step of 0.1 minutes as this was found to be the minimum time necessary to achieve convergence of the model output.  We calculate the new center position and remaining grid points using the calculated nose position and CME size. 

We assume that $\vec{v}_{nr}$ remains in the non-radial direction.  As the CME deflects, the non-radial direction changes and $\vec{v}_{nr}$ changes accordingly.  As the CME propagates out radially, angular momentum should be conserved in the absence of any deflection forces.  ForeCAT determines the CME's angular momentum, $L=M_{CME} v_{nr} R$, which remains constant between consecutive time steps before the effects of the deflection forces are included.  Setting $L(t+\Delta t) = L(t)$ yields 
\begin{equation}
v_{nr}(t+\Delta t) = v_{nr}(t) \frac{R(t)}{R(t+\Delta t)}
\end{equation}
.  The full change in $v_{nr}$ includes the acceleration term from the deflection forces. 
\begin{equation}\label{eq:vangmom}
\vec{v}_{nr}(t+ \Delta t) = \vec{v}_{nr}(t) \frac{R(t)}{R(t+\Delta t)} + \vec{a}(t) \Delta t
\end{equation}

When neither drag nor deflection forces can noticeably influence the CME's motion, it will continue to deflect through interplanetary space with constant angular momentum, $L$.  Equation \ref{eq:vangmom} shows that the deflection velocity, $v_{nr}$ decreases as 1/R when angular momentum is conserved. This velocity can be converted to an angular velocity $\omega = v_{nr} / R$.  Expressing the angular velocity in terms of the time derivative of an angular postion, $\theta$, we find

\begin{equation} \label{eq:vrr}
\omega = \frac{d\theta}{dt} = v_{nr,0} \frac{R_0}{R^2}
\end{equation} 

where $v_{nr,0}$ is the deflection velocity at the time the angular momentum stops changing, which occurs at some radial distance, $R_0$.  The time derivative can be converted into a radial derivative using the CME's radial velocity, $v_{CME}$, if we assume constant radial propagation.
\begin{equation}
dR = v_{CME} dt
\end{equation}
This yields
\begin{equation}
\frac{d\theta}{dR} = \frac{v_{nr,0}}{v_{CME}} \frac{R_0}{R^2}
\end{equation}
which integrates to
\begin{equation}\label{eq:thetaangmom}
\theta(R) = \theta_0 + \frac{v_{nr,0}}{v_{CME}} (\frac{1}{R_0} - \frac{1}{R})
\end{equation}
where $\theta_0$ is the angular position at $R_0$.  Both the latitudinal and longitudinal motion should behave as Eq. \ref{eq:thetaangmom} when angular momentum is conserved.  This equation should be used as test whether observed CME's are actively being accelerated versus simply propagating with the angular momentum obtained in the low corona.

\subsection{CME Rotation}
In a similar way to forces causing deflection through a linear acceleration, a torque, $\tau$, causes rotation via an angular acceleration, $\alpha$, which changes the angular momentum, $L$.  The torque is defined as the cross product of the lever arm, $r$, and the force, $F$.   
\begin{equation}
\tau = r \times F = I \alpha = \frac{dL}{dt}
\end{equation}
Appendix \ref{MoI} presents a derivation of the moment of inertia, $I$ of the CME torus.  Since we calculate the deflection force at multiple locations along the toroidal axis, we can determine the torque on the CME torus that will cause the CME to rotate about the axis pointing from the center of the Sun through the normal direction at the nose of the CME.  This causes a change in the tilt of the CME.  At each time step, we determine the change in the angular momentum of the CME based on the torques.  The rotation rate, $\omega$ is then determined using the moment of inertia.
\begin{equation}
\omega (t + \Delta t) = \frac{L(t) + \tau(t) * \Delta t}{I(t)} 
\end{equation}
Since we conserve angular momentum, the rotation rate can change due to either a torque, or a change in the moment of inertia.  As the CME expands the moment of inertia increases and the torque becomes negligible as the magnetic forces decay with distance.  This causes the CME rotation to cease at large distances.

\section{Solar Background}\label{sw}

The background solar wind through which the CME propagates determines the forces that influence the CME during propagation to 1 AU.  ForeCAT's models for deflection and drag depend on the background solar wind velocity and density, as well as the solar magnetic field.  Both the angular and radial dependencies of these parameters must be described by the solar models coupled with ForeCAT.

ForeCAT's background solar wind density model utilizes the results of \citet{Guh06}, hereafter G06, which allows us to better describe the three dimensional effects due to the location of the CHs and SB.  The K13 version of ForeCAT used the one-dimensional density model of \citet{Che96}.  In addition to the spatial variations, the G06 and \citet{Che96} density models vary by nearly an order of magnitude in the low corona.  We consider the effect of the change in the density model in appendix \ref{FC1comp}.

G06 describes the solar wind density in two dimensions as
\begin{equation}\label{eq:G06a}
N(R, \theta) = N_p(R) + [N_{cs}(R) - N_p(R)] \mathrm{e}^{-\lambda^2 / w^2}
\end{equation}
where $\theta$ is the latitude, $R$ is the radial distance, $\lambda$ is the angular distance from the HCS, and $w$ is an angular width describing the range of the influence of the HCS on the density, set equal to 34$\mydeg$ in G06.  $N_p$ and $N_{cs}$, which respectively describe the radial evolution of the polar and current sheet density, are of the form
\begin{equation}\label{eq:G06b}
N_x = a_1 \mathrm{e}^{a_2 z + a_6 z^2}z^2[1+a_3 z + a_4 z^2 + a_5 z^3]
\end{equation}
where $z$ equals $1/R$.  At any given distance this form produces nearly constant density in the poles and smooth variation toward the current sheet.  G06 determine the polynomial coefficients from a combination of white light coronagraph images from both Mauna Loa, Hawaii and from the Solar and Heliospheric Observatory Large Angle and Spectrometric Coronagraph Experiment (SOHO/LASCO)  in the range 1.16-30$\rsun$.  In addition, Ulysses in situ plasma measurements were extrapolated to 1 AU and provided additional constraints.  The data above 34$\mydeg$ determine the polar coefficients and the data between 5$\mydeg$ and 34$\mydeg$ determined the current sheet coefficients.  This model works best for solar minimum configuration.

We make several modifications to the two-dimensional G06 model.  Instead of restricting $\theta$ and $\lambda$ to latitudinal distances, we consider both latitude and longitude.  This allows for description of the solar wind density in three dimensions.  In addition, we vary the width of the current sheet region close to the Sun, similar to the earlier work, \citet{Guh96}.  

\begin{figure}
\includegraphics[scale=0.3]{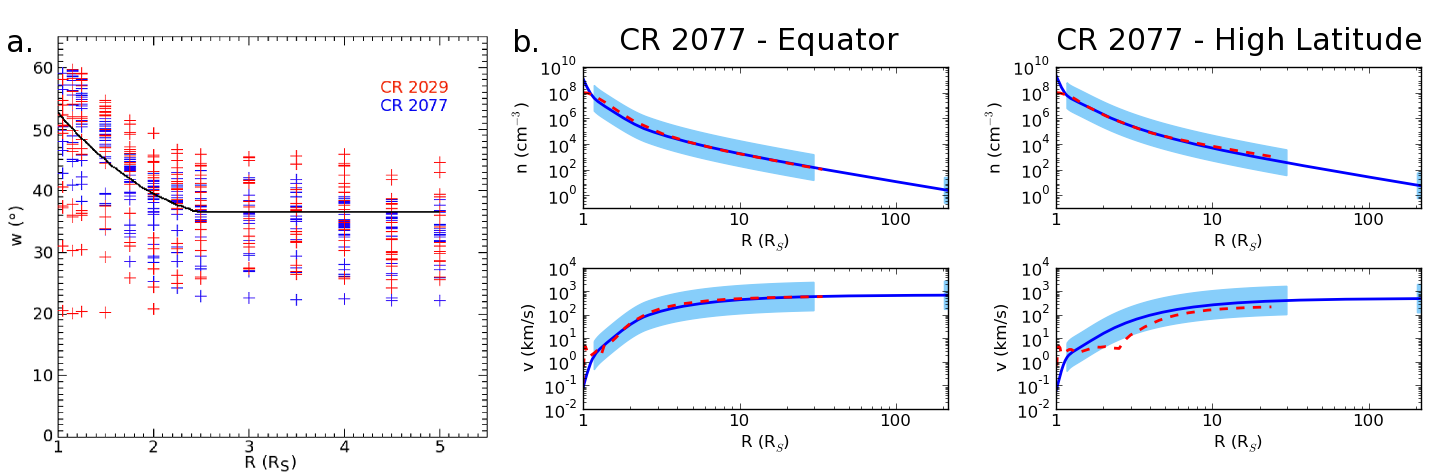}
\caption{Fig. \ref{fig:wfit}(a) shows fits of Gaussian width versus radial distance.  Each point shows the fit for a 10$\mydeg$ bin of longitude for either CR 2029 (red) or CR 2077 (blue).  The black line shows the quadratic best fit to the data below 2.5 $\rsun$. A constant value of $w$ is assumed above this height. The blue line in Fig. \ref{fig:wfit}(b) shows ForeCAT's solar number density (top) and velocity (bottom) versus distance for the equator (left) and high latitude (right) of CR 2077.  The red line shows the results from the MHD simulation and the blue shaded regions indicates the distances which data are used to determine the \citet{Guh06} density model.}\label{fig:wfit}
\end{figure}

To determine the width we use the results of a 3D MHD simulation that produces a steady state solar wind output of an Alfv\'en wave driven solar wind including the effects of surface Alfv\'en wave damping \citep{Eva12}.  We run separate simulations for each Carrington Rotation used.  To determine the width, we group the MHD results into 10$\mydeg$ bins of longitude and fit a Gaussian of the form $\mathrm{e}^{-\lambda^2 / w^2}$ to each bin.  Figure \ref{fig:wfit} shows the Gaussian width for each longitude bin for CR 2029 (red) and CR 2077 (blue) versus distance.  Below 2.5 $\rsun$, a quadratic polynomial describing the variation of the width with distance is determined from the measurements from all bins.  The data show significant scatter around this fit, however no systematic difference occurs between the two CRs.  Above 2.5 $\rsun$ a constant value is assumed.  The top panels of Figure \ref{fig:wfit}(b) compare the ForeCAT density model (blue lines) with the MHD results at the equator (left) and high latitude (right).  We find good agreement for both cases.

The ForeCAT solar wind velocity model assumes a purely radial solar wind outflow, a reasonable assumption in the open solar corona. We assume a constant mass flux and determine the solar wind velocity using the radial distance and the density.  Similar to the G06 density model which determines a balance between a polar CH and current sheet component, we use two different fluxes $\dot{M}_{p}=$1.19x10$^{-14}$ $M_{\Sun}$ yr$^{-1}$ and $\dot{M}_{cs}=$1.87x10$^{-14}$ $M_{\Sun}$ yr$^{-1}$.  The velocity is then

\begin{equation}
v_{SW} = \frac{\dot{M}_{p} + [\dot{M}_{cs} - \dot{M}_{p}] \mathrm{e}^{-\lambda^2 / w^2}}{m_p N R^2}
\end{equation}  
where $N$, $\lambda$, and $w$ are the same as in Eq. \ref{eq:G06a}, and $m_p$ is the proton mass. The bottom panels of Fig. \ref{fig:wfit}(b) compare the ForeCAT velocity with the MHD results.  We find good agreement for the equatorial velocity at all distances and the high latitude velocity beyond about 5 $\rsun$.

The non-radial drag force does not depend on the magnitude of the radial solar wind velocity so our chosen velocity model only affects the determination of the interplanetary magnetic field, which we describe below.

ForeCAT utilizes a PFSS model to describe the solar magnetic field.  The PFSS model assumes a current free configuration below the source surface so that the magnetic field can be described as the gradient of a potential and then the magnetic field can be calculated using a sum of Legendre polynomials \citep{Alt69, Sch69, Alt77}.  \citet{Ril06a} found that the magnetic structures of MHD models and PFSS models often closely match on global scales.

For both PFSS and MHD magnetic field solutions, the magnetic field strength decreases rapidly with distance.  Observations of type II radio burst suggest that these models may produces a magnetic field that decays too rapidly with distance \citep{Man03}.  \citet{Man03} show the formation, decay, and reformation of a Type II radio burst below 6 $\rsun$, suggesting there should be a local maximum in the profile of the Alf\'en speed versus radial distance.  \citet{Eva08} shows that MHD models often do not reproduce an Alfv\'en profile with the local minimum and maximum.  In section \ref{scaleB}, we explore the sensitivity of ForeCAT deflections to artificially scaling the PFSS magnetic field so that it decreases less rapidly with distance.

Beyond the source surface we use the Parker interplanetary magnetic field model to incorporate the effects of solar rotation on the background magnetic field.  Beyond the source surface, at a distance $R$ the magnetic field is defined as

\begin{equation}\label{eq:PSr}
B_r = B_{SS} \left(\frac{R_{SS}}{R}\right)^2
\end{equation}

and

\begin{equation}\label{eq:PSp}
B_{\phi} = B_{SS} \left(\frac{R_{SS}}{R}\right)^2 (R - R_{SS}) \frac{\Omega_{\Sun} sin \theta}{v_{SW}}
\end{equation}
where $B_{SS}$ is the magnitude of the magnetic field at the source surface,  $R_{SS}$ is the source surface distance, $\Omega_{\Sun}$ is the solar rotation rate, and $\theta$ is the colatitude.

\begin{figure}
\includegraphics[scale=0.75]{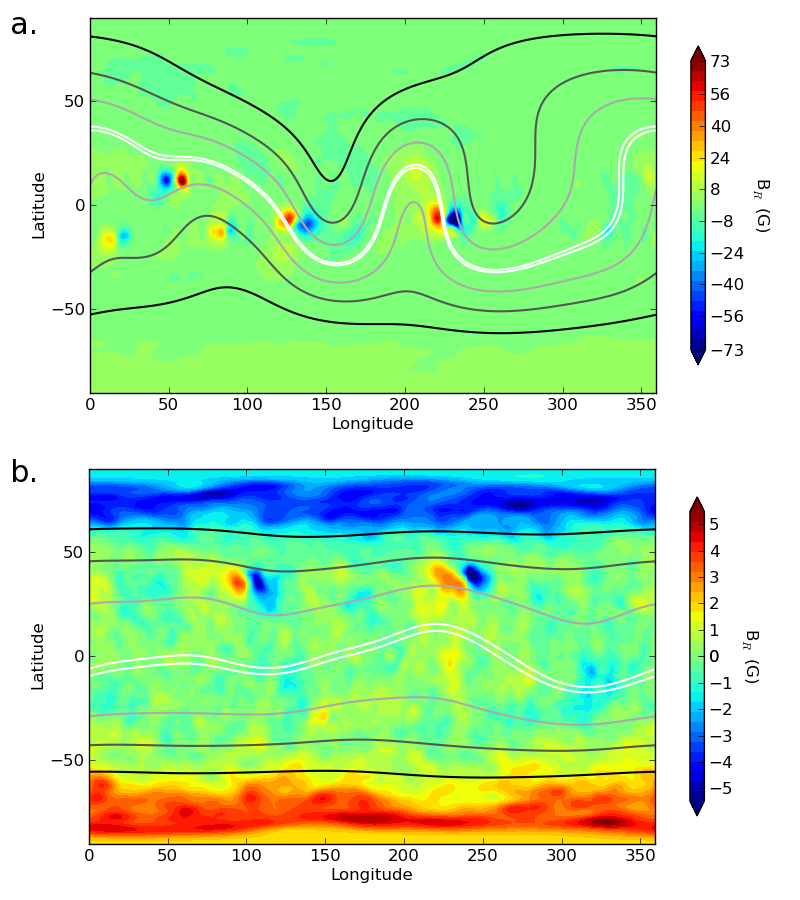}
\caption{The panels show the solar magnetic field for CR 2029 (\ref{fig:CRs}(a)) and CR 2077 (\ref{fig:CRs}(b)).  The color contours represent the radial magnetic field at 1.05 $\rsun$, which show the location of the ARs.  The line contours represent the magnetic field strength at the source surface height, 2.5 $\rsun$, with white contours indicating the weakest magnetic field strength, which occurs near the HCS.}\label{fig:CRs}
\end{figure}

Here, we consider two different CR rotations: CR 2029 (April-May 2005), a declining phase background, which we used in K13, and CR 2077 (November-December 2008), a solar minimum background.  Figure \ref{fig:CRs} shows the magnetic field for CR 2029 (top) and CR 2077 (bottom). The white lines show the location of the magnetic minimum, which corresponds to the HCS on global scales.  The location of the HCS does not vary with respect to Carrington longitude beyond 2.5 $\rsun$, however solar rotation will cause a CME to drift toward smaller Carrington longitudes.

Fig. \ref{fig:CRs} highlights the fundamental differences between a declining phase and a solar minimum CR.  First, the HCS remains relatively flat during solar minimum, but is more inclined during the declining phase.  Second, the ARs are located at high latitudes during solar minimum, but migrate toward the equator during the declining phase.  Additionally, there are fewer ARs during solar minimum.  Finally, the background magnetic field is significantly weaker during solar minimum than in the declining phase.  The AR magnetic field strength greatly increases in the declining phase, as well as the quiet sun magnetic field strength, to a lesser extent.  All of these factors directly impact the ForeCAT deflections.

\section{ForeCAT Deflections}\label{FCdefs}
ForeCAT CMEs are initiated using three initial position parameters (latitude, longitude, and tilt), the four initial shape parameters ($a$, $b$, $c$, and $d$ shown in Fig. \ref{fig:diag}), and the CME mass and final radial propagation velocity.  Table 1 lists CME input parameters for the CMEs considered in this work: the initial latitude, longitude, tilt, and length $c$ of the CME torus.  The number in the CME name corresponds to the CR.  An upper case letter in the CME name indicates a CME initiated at a PIL associated with ARs whereas lower case letters are at initiated in the quiet sun.  Typically, observations can constrain all of the input parameters for individual cases, except for the CME mass.  In this work, unless otherwise specified, CMEs are initiated with a mass of 10$^{15}$ g, final propagation velocity of 475 km s$^{-1}$, and cross-sectional width of 0.01 $\rsun$ at a distance, $d$, of 1.05 $\rsun$.  The shape parameter $a$ is set equal to $c$ for all cases and the three phase radial propagation model has a $v_g=$80 km s$^{-1}$ and transition radii at $r_{ga}=$1.5 and $r_{ap}=$3.0 $\rsun$. We do not include CME rotation unless explicitly stated.

\begin{table}
\newcolumntype{d}[1]{D{.}{.}{#1}}
\begin{tabular}{|l|d{3}|d{3}|d{3}|d{3}|}
\hline
CME               & \mathrm{Latitude} \; (\mydeg) & \mathrm{Longitude}  \; (\mydeg) & \mathrm{Tilt}  \; (\mydeg) & \mathrm{Width (c)}  \; (\rsun)\\
\hline
2029A           & -15.40 &  17.00 & -72.0 & 0.196 \\
2029B           &  12.40 &  53.25 &   90.0 & 0.128 \\
2029C           & -12.60 &  86.65 & -86.0 & 0.139 \\
2029D           &  -7.90 & 131.90 &  48.0 & 0.147 \\
2029E           &  -5.40 & 226.00 &  55.0 & 0.240 \\
2029F           &  -5.10 & 256.20 & -44.0 & 0.166 \\
2029G           & -16.00 & 321.85 &  65.0 & 0.139 \\
2029H           &  17.20 & 348.40 & -54.0 & 0.142 \\
2029a           & -45.60 &  36.70 & -13.0 & 0.169 \\
2029b           &  37.20 & 121.90 & -31.9 & 0.275 \\
2029c           & -40.90 & 236.10 &  26.0 & 0.207 \\
2029d           &  33.40 & 305.30 & -24.0 & 0.207 \\
2077A           &  36.80 &  99.50 & -54.0 & 0.150 \\
2077B           &  35.00 & 238.30 & -35.0 & 0.250 \\
2077a           &  44.30 &  48.50 & -10.0 & 0.219 \\
2077b           & -34.50 &  67.70 & -82.0 & 0.235 \\
2077c           & -20.00 & 267.45 &   6.0 & 0.275 \\
2077d           &  38.60 & 289.20 &  67.0 & 0.292 \\

\hline
\end{tabular}
\caption{Initial parameters for the CMEs in this paper.  The CME height is set equal to the width and the cross-sectional width set at one-fourth of the width.}
\end{table}

\begin{figure}
\includegraphics[scale=0.4]{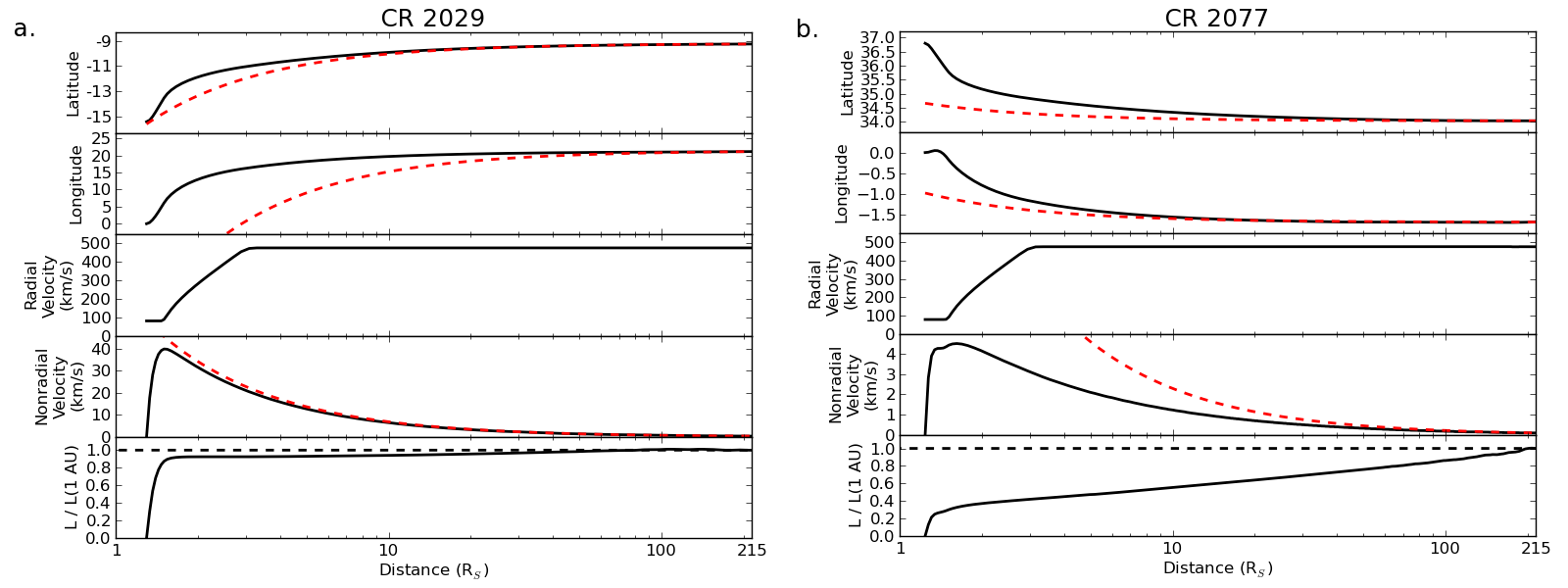}
\caption{Deflection results for CME 2029A (panel a) and CME 2077A out to 1 AU (panel b). From top to bottom, the panels show the CME's latitude, longitude, radial velocity, nonradial velocity, and the ratio of the angular momentum relative to the angular momentum at 1 AU (indicated with a dashed black line).  The dashed red lines indicate the behavior expected for a CME propagating with constant angular momentum.}\label{fig:CPAcontrols}
\end{figure}

The left panel of Figure \ref{fig:CPAcontrols}(a) shows ForeCAT results for the trajectory of CME 2029A out to 1 AU.  The top two plots show the change in the CME's latitude and longitude versus radial distance.  We determine the CME's longitude in a fixed observer coordinate system, similar to the Earth-centric Stonyhurst coordinates, but with the initial longitude of the CME set to the initial longitude of the CME in Carrington coordinates.  As we are not comparing with specific observations, we do not account for orbital motion of the fixed observer, as for a spacecraft orbiting at 1 AU, which can cause 4$\mydeg$ apparent eastward motion for a CME that takes four days to propagate to 1 AU.  The longitudinal motion is then solely the result of the CME deflection.  The next two panels show the radial and non radial velocity of the CME.  The bottom panel shows the fraction of the angular momentum relative to the total angular momentum at 1 AU.  The dashed red lines in the latitude, longitude, and nonradial velocity panels correspond to deflection with constant angular momentum.   Equations \ref{eq:vrr} and \ref{eq:thetaangmom} are fit to the ForeCAT deflection near 1 AU.  Deviation from the dashed red lines indicates a noticeable difference in the angular momentum from the 1 AU values and that the deflection forces are still influencing the CME motion at that distance.

It can be seen that the strong magnetic forces in the low corona cause a strong non-radial acceleration and cause the CME to begin deflecting in both latitude and longitude.  The non-radial velocity quickly increases close to the Sun and reaches a maximum value by 1.5 $\rsun$.  This maximum non-radial speed is a fraction of the CME's radial velocity.  The angular momentum initially also rapidly increases until the CME reaches 1.5 $\rsun$.  It continues to increase beyond this distance, but at a significantly slower rate.  The continued increase in the angular momentum is sufficiently slow that the trajectory and velocity profiles between 1.5 and 5 $\rsun$ are well reproduced by the analytic equations describing deflection from constant angular momentum.  We find that this continued gentle increase in the angular momentum persists out to 1 AU.  This causes a slight deviation in the behavior between 1.5 $\rsun$ and 1 AU from Eq. \ref{eq:thetaangmom} for any single choice of $R_0$, $\theta_0$, and $v_{nr,0}$, however, the equation remains valid on shorted distance scales.  This continued gradual increase in the angular momentum will be further explored in a future work.

A CME will continue to deflect in interplanetary space unless some force can halt the deflection motion.  The continued increase in the angular momentum of this CME can only result from the magnetic forces accelerating (rather than decelerating) the CME, so drag is the only force which can potentially decelerate the CME.  For other CMEs, the magnetic forces may change directions and could contribute to the deceleration.   

This CME deflects toward the HCS but does not reach it by 1 AU.  For this CME, the majority of the deflection occurs below 10 Rs, however, some motion continues during further propagation to 1 AU.  Beyond 10 $\rsun$, this CME deflects an additional 0.7$\mydeg$ and 1.5$\mydeg$ in latitude and longitude, or 11\% and 7\% of the respective total deflections.

Fig. \ref{fig:CPAcontrols}(b) shows the same as Fig. \ref{fig:CPAcontrols}(a) but for CME 2077A.  This CME deflects primarily in latitude, which corresponds to deflection toward the HCS during solar minimum.  Much of the deflection behavior is similar to that of CME 2029A, however the decrease in the magnetic field strength and weaker magnetic gradients present at solar minimum cause a decrease in the total amount of deflection as compared to the declining phase case.  The angular momentum has a much smaller initial increase than seen for CME 2029A, which causes the continued gradual rise in the angular momentum to be more noticeable in Fig. \ref{fig:CPAcontrols}(b).  Again the majority of the deflection occurs below 10 $\rsun$ with 11\% and 7\% of the total latitudinal and longitudinal deflection occurring beyond 10 $\rsun$.

For both CME 2029A and 2077A the latitudinal deflection brings the CME closer to the solar equator throughout the propagation to 1 AU.  CME 2029A experiences a consistent westward deflection, which results from a combination of both local and global magnetic gradients.  ForeCAT CME's are initiated in approximate local equilibrium by aligning the CME torus with the AR PIL, as viewed in the photosphere.  As the CME begins to rise, the local gradients change and can cause the CME to deflect in a direction differing from that of the global magnetic gradients that deflect CMEs toward the HCS.  The local magnetic gradients decay rapidly with height so as the CME continues to rise the global gradient begin to dominate and can cause a change in the direction of deflection.  The strongest local gradients result from imbalances in the opposite polarity flux systems of the AR and cause deflection perpendicular to the PIL, which tends to correspond to the longitudinal direction.  For CME 2029A both local and global gradients cause westward motion so we do not see a change in the direction of deflection.  For CME 2077A the longitudinal deflection is negligible due to the flat HCS and weak local magnetic gradients present during solar minimum, however we do see a slight change in the direction.

\section{Sensitivity of the Deflection to ForeCAT Model Input Parameters}\label{modelsens}
Here we explore the sensitivity of the results to the input parameters used in the radial propagation model as well as the magnetic field model.

\subsection{Radial Propagation}\label{propsens}
We explore the effects of the chosen radial propagation model parameters on ForeCAT results.  These parameters affect the speed of the CME in the corona as here we explore the effects specific to this location.  If the CME spends more time close to the Sun in regions of strong magnetic field, the deflection forces act upon the CME longer, increasing the non-radial speed and therefore increasing the deflection.  As shown in section \ref{expdep} the velocity also affects the amount of time a CME spends deflection before it reaches 1 AU, however we consider this effect when we explore the sensitivity to CME input parameters in section \ref{MvV}

\begin{figure}
\includegraphics[scale=0.3]{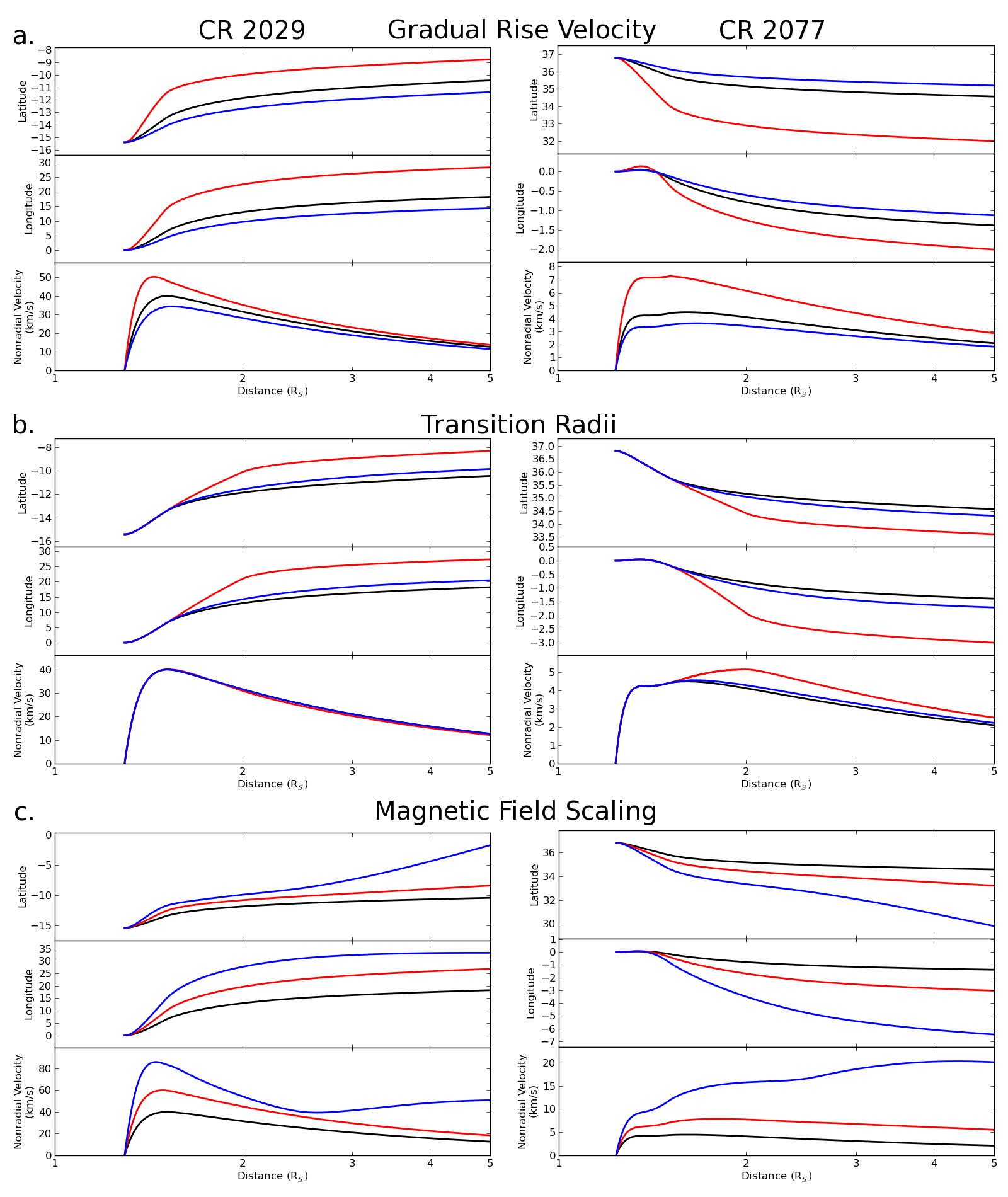}
\caption{Variations in the deflection resulting from changes to initial radial propagation speed, $v_g$, of the three-phase propagation model (Fig. \ref{fig:RVB}(a)), the distances used for the transition between the gradual rise and acceleration phase, $r_{ga}$, and the transition between the acceleration and constant propagation phase, $r_{pa}$ (Fig. \ref{fig:RVB}(b))), and the magnetic field model (Fig. \ref{fig:RVB}(c)).  All panels have the same format as Fig. \ref{fig:CPAcontrols} but only show the latitude, longitude, and nonradial velocity.  The left panels show results for CME 2029A and the right panels show results for CME 2077A.}\label{fig:RVB}
\end{figure}

Figure \ref{fig:RVB}(a) shows the effect of varying the gradual rise velocity, $v_g$, defined in section \ref{PED} as the initial radial velocity until the CME reaches the distance $r_{ga}$.  Fig. \ref{fig:RVB}(a) show results for CME 2029A (left) and CME 2077A (right) for distances below 5 $\rsun$ to highlight the effect in the low corona.  The black lines correspond to the control cases shown in Fig. \ref{fig:CPAcontrols} that have $v_g$ equal to 80 km s$^{-1}$.  The other cases have all of the same input parameters as the control but with $v_g$ increased to 100 km s$^{-1}$ (blue lines) or decreased to 50 km s$^{-1}$ (red lines).  For both CRs, when the gradual rise velocity decreases we see the expected increase in the non-radial velocity.  Since these CMEs have the same final radial propagation speed, the increase in the non-radial velocity causes an increase in the deflection.  Decreasing $v_g$ causes increases of 1.6$\mydeg$ and 10.2$\mydeg$ in the latitudinal and longitudinal deflection for CR 2029 at 1 AU and 2.7$\mydeg$ and 0.7$\mydeg$ for CR 2077.  By comparing the percent increase in the latitudinal and longitudinal directions, respectively 26\% and 48\% for CR 2029 and 99\% and 43\% for CR 2077, we find that varying $v_g$ does cause a slight change in the direction of the CME deflection.  Decreasing the gradual rise velocity increases the time spent in the low corona, increasing the effect that local gradients can have on a CME's deflection.

We also determine the effects from having fixed distances $r_{ga}$ and $r_{ap}$ determine the the transition between the gradual rise and acceleration phase and the transition between the acceleration and constant propagation phase (Figure \ref{fig:RVB}(b)).  The CMEs behave the same during the initial gradual rise phase (below 1.5 $\rsun$).  When $r_{ga}$ is increased to 2.0 $\rsun$ (red lines) the CME does not begin accelerating until farther distances allowing it more time to be accelerated by the strong deflection forces in the low corona.  This causes a small increase in the non-radial velocity that leads to an increase in deflection.  Increasing the length of the acceleration phase (blue line) causes the CME to move slower throughout the duration of the acceleration phase.  This causes negligible increases in the deflection velocity and the amount of deflection during the acceleration phase.  Within the constant propagation phase the CME continues to deflect at nearly the same rate as the control case.

\subsection{Background Magnetic Field}\label{scaleB}
We consider the effects of the background magnetic field on the CME deflection.  As described in section \ref{sw}, observations of some type II radio bursts suggest that the PFSS magnetic field may decrease too rapidly with distance in the low corona \citep{Man03}.  We artificially increase the PFSS magnetic field by a factor of $R$ or $R^2$ below the source surface height (2.5 $\rsun$).  Above the source surface we replace the $B_{SS}$ in equations \ref{eq:PSr} and \ref{eq:PSp} with the source surface value obtained from the scaled model. These artificially scaled magnetic field models do not represent physical solutions, however, they allow us to explore ForeCAT's sensitivity to the background magnetic field.

Fig. \ref{fig:RVB}(c) shows the resulting deflections after increasing the magnetic field by a factor of $R$ (red line) or $R^2$ (blue line).  For both CRs, we see that when the magnetic field decreases less quickly with distance, the non-radial velocity increases.  For CR 2077, the slower decaying magnetic field also causes noticeable increases in the non-radial velocity out to further distances.  For CR 2029, the non-radial velocity still reaches a maximum value by 1.5 $\rsun$.  These results show ForeCAT is quite sensitive to the background magnetic field in the low corona.  By comparing observed CME deflections with ForeCAT results we may be able to constrain the rate at which the coronal magnetic field decreases with distance and make predictions about the in situ observations Solar Probe will make at its closest approach to the Sun.

\subsection{Drag Model Variations}\label{dragvar}
\begin{figure}
\includegraphics[scale=0.4]{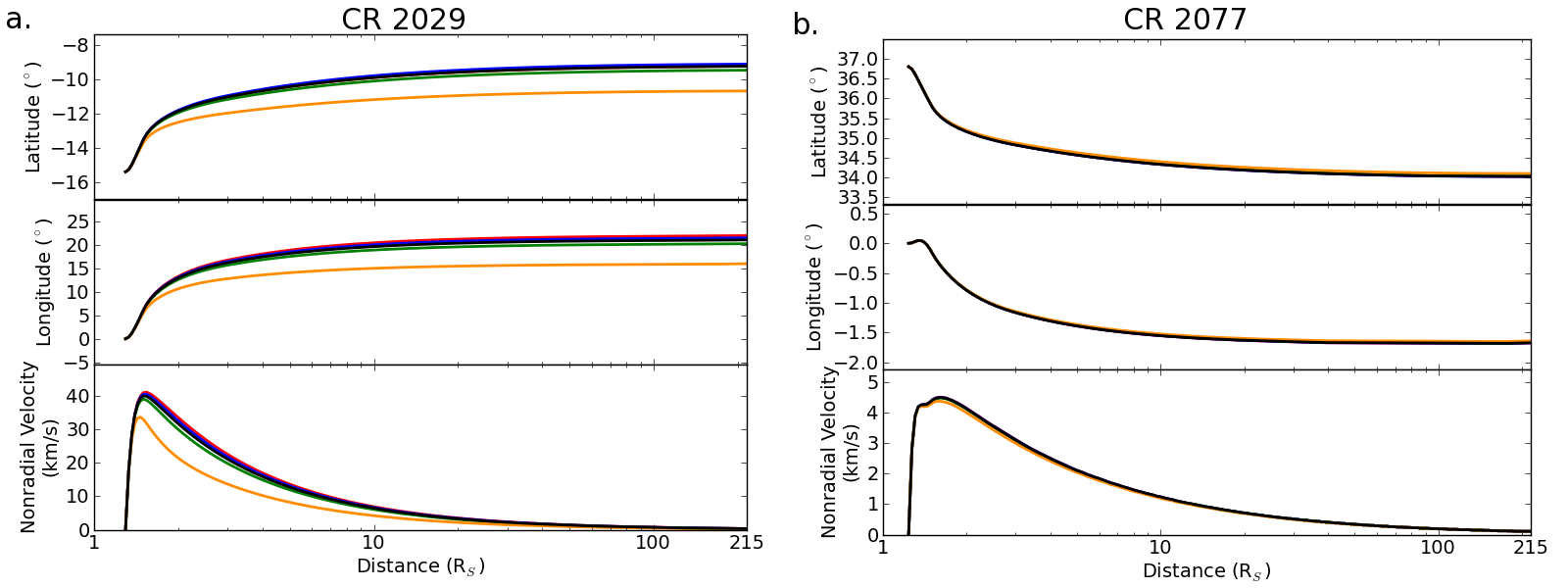}
\caption{Variations in CME deflection resulting from decreasing the drag coefficient to 0.5 (blue line) or zero (red line) or increasing the drag coefficient to 2.0 (green line) or 10.0 (yellow line).}\label{fig:CPACd}
\end{figure}

The drag coefficient, $C_d$ is typically assumed to have some value near unity.  Figure \ref{fig:CPACd} shows that the deflection does not depend strongly on the drag coefficient unless values larger than typically used are considered.  CME 2077A shows no visible difference between the lower (red and blue lines) and higher (green line) drag cases.  The changes are more noticeable due to the larger deflection speed of CME 2029A.  For CME 2029A, the low drag case shows more deflection, as expected from the scaling of Eq. \ref{scalingeq}, however the difference is negligible.  When the drag coefficient is increase to ten (orange lines), we begin to see a noticeable decrease in the deflection for both CRs. Even for this extreme case, the drag is unable to stop the CMEs deflection, and the CME continues deflecting through interplanetary space.

While coronagraph observations have clearly shown the effects of drag in the radial direction \citep{Mal10, Tem11, Iju13}, we find very little effect in the deflection.  Although the solar wind density is high close to the Sun, the deflection velocity is small, typically 10-50 km s$^{-1}$.  Only with $C_d$ as large as 10 do we find a non-negligible effect from drag, and only for the rapidly deflecting CME in the strong magnetic background. 

\subsection{CME Input Parameters}\label{CMEinputs}
Here we assume the propagation model used in the control cases presented in \ref{FCdefs} (gradual rise velocity of 80 km s$^{-1}$ and transition distances of 1.5 and 3.0 $\rsun$) and explore how the deflection changes for different CME properties such as mass, final propagation velocity, and shape.  Observed CMEs exhibit significant variations in both mass and velocity \citep{Gop09}.  The initial cross-sectional width of a flux rope is also highly uncertain.  Exploring the range of observed parameters produces a range of expected deflections for any initial location within a CR.

\subsubsection{Mass}\label{Mass}
\begin{figure}
\includegraphics[scale=0.3]{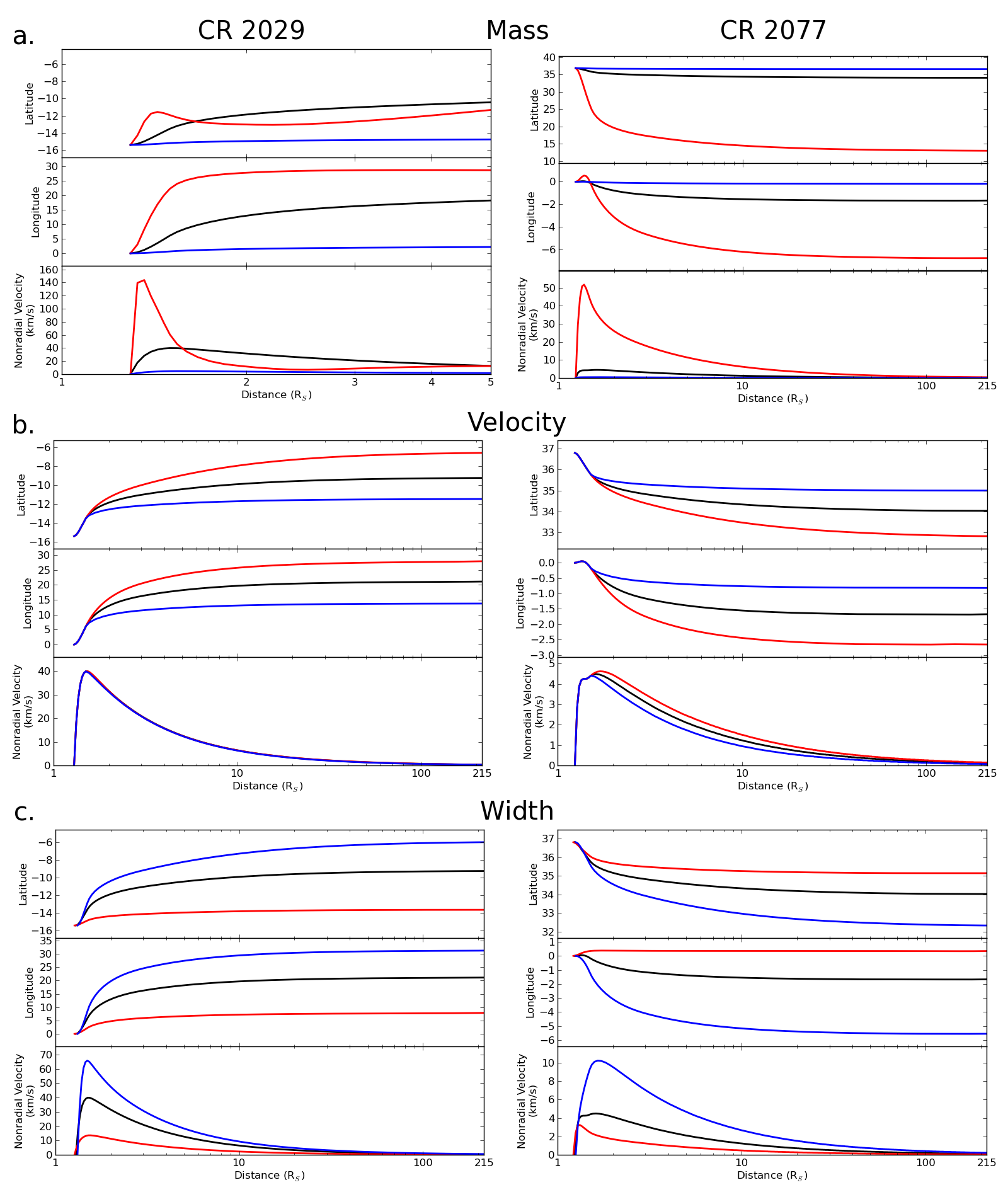}
\caption{Similar to Fig. \ref{fig:RVB} but for variations in CME deflection resulting from changes to the CME mass (Fig. \ref{fig:MVW}(a)), final propagation velocity (Fig. \ref{fig:MVW}(b)), and cross-sectional width (Fig. \ref{fig:MVW}(c)).}\label{fig:MVW}
\end{figure}

The SOHO/LASCO CME catalog \citep{Gop09} contains CMEs with masses between 10$^{13}$ and a few times 10$^{16}$ g.  Figure \ref{fig:MVW}(a) shows ForeCAT results for low mass (10$^{14}$, red line) and high mass (10$^{16}$, blue line) versions of CME 2029A (left) and 2077A (right).  We do not consider masses as low as 10$^{13}$ g as the lowest mass CMEs tend to be less important for space weather effects.  The format of Fig. \ref{fig:MVW} is the same as the left panels of Fig. \ref{fig:CPAcontrols}, however we only show the deflection velocity for each case.

For both CRs, the less massive CMEs obtain faster non-radial velocities.  The decrease in density causes larger accelerations from the same deflection forces leading to larger deflections.  This matches the trend expected from section \ref{expdep}.  For CR 2029, the low mass CME shows more effects from the local gradients with a brief reversal in the latitudinal motion below 2 $\rsun$.  This low mass CME then returns to deflecting in a similar manner to its more massive counterparts. The low mass CME also reverses its longitudinal motion near the same distance as the latitudinal reversal.  This causes the low mass CME to gradually move back towards its initial longitude.  For CR 2077, the low mass CME initially moves in the longitude direction opposite that of the more massive CMEs due to a weak local gradient.  This motion quickly reverses and the low mass CME begins deflecting in the same direction as the other CMEs.

\subsubsection{Final Propagation Speed}\label{Vel}
We next consider the variation in the final deflection due to different CME propagation velocities, $v_{CME}$.  The control cases have a final propagation velocity of 475 km s$^{-1}$ during the constant propagation phase (above 3 $\rsun$).  The coordinated data analysis workshop (CDAW) catalog \citep{Yas04, Yas08, Gop09} contains CMEs observed with LASCO.  CDAW CMEs which occurred between 1996 and 2006 have an average plane-of-sky velocity of 475 km s$^{-1}$ \citep{Gop09}.  However, CMEs can have propagation velocities up to several thousand km s$^{-1}$, with faster CMEs tending to be more geoeffective \citep{Ric10}.  Figure \ref{fig:MVW}(b) shows results for CMEs 2029A (left) and 2077A (right) with different constant radial propagation speeds.  The blue lines show cases with final radial propagation speeds of 1000 km s$^{-1}$, and the red lines show cases with final radial propagation speeds of 300 km s$^{-1}$.  The cases have the same radial propagation and non-radial speeds during the slow rise phase (below 1.5 $\rsun$) but the trajectories and speeds begin to differ when the CME begins accelerating, as shown in Fig. \ref{fig:MVW}(b).  As expected from section \ref{expdep}, the deflection increases as the radial propagation speed decreases.  We see less of a dependence on speed than we do for mass, which conflicts the analysis of section \ref{expdep}.  However, for those scalings we assume that constant forces cause the deflection when the forces actually rapidly decrease with distance.  By the time the CME has reached the final propagation speed the deflection forces no longer accelerate the CME in the non-radial direction.  This implies that the deflection will be proportional to $v_{CME}$, as in Eq. \ref{eq:thetaangmom}, as opposed to $v^2_{CME}$ since the CME has a constant non-radial velocity.

\subsubsection{CME Shape Parameters}\label{shape_vars}
The CME width, $c$, in Fig. \ref{fig:diag} is determined by the length of the PIL.  Both observations \citep{Tri09, She12, Tri12} and simulations \citep{Gib06} show that partial filament eruptions can occur.  Whether a full or partial eruption occurs is still an area of open research, so for simplicity we assume the entire length of the PIL contributes in forming the flux rope of the erupting CME.  We have made the assumption that the initial height of the flux rope equals the half-width ($a=c$).  The actual initial configuration of individual CMEs is certainly more complex than this and will be accounted for in future works when we compare ForeCAT results to specific observations.  This leaves only the cross-sectional radius, $b$, as a free parameter.  For the control cases we have assumed $b=$0.25$c$.  To explore ForeCAT's sensitivity to this parameter we consider ``wide'' and  ``narrow'' cases with the cross-sectional radii multiplied by a factor of 1.5 or 0.5, respectively the blue and red lines in Fig. \ref{fig:MVW}(c).

For both CRs, we find that the wide CMEs tend to have faster non-radial velocities and deflect more than the narrow CMEs.  As shown in section \ref{expdep} an increase in the width will decrease the tension and drag forces but it also decreases the density.  The net result is an increase in the acceleration and resulting deflection.  For CME 2029A, we see that changing the width can cause the deflection to vary by 10$\mydeg$ in both the latitudinal and longitudinal deflections.  The effect is less noticeable, of order a few degrees, for the smaller deflections of CME 2077A.

\subsubsection{Mass and Velocity}\label{MvV}
Results for individual ForeCAT CMEs show that variations in the initial CME parameters can have a significant effect on the CME deflection.  Due to the simplicity of ForeCAT, we can easily constrain the extent of the deflections for any initial CME location by sampling a large number of CMEs covering a wide range in parameter space.  Here we further explore the sensitivity to mass and velocity as these parameters are better constrained than the initial cross-sectional width of a flux rope.  

\begin{figure}
\includegraphics[scale=0.4]{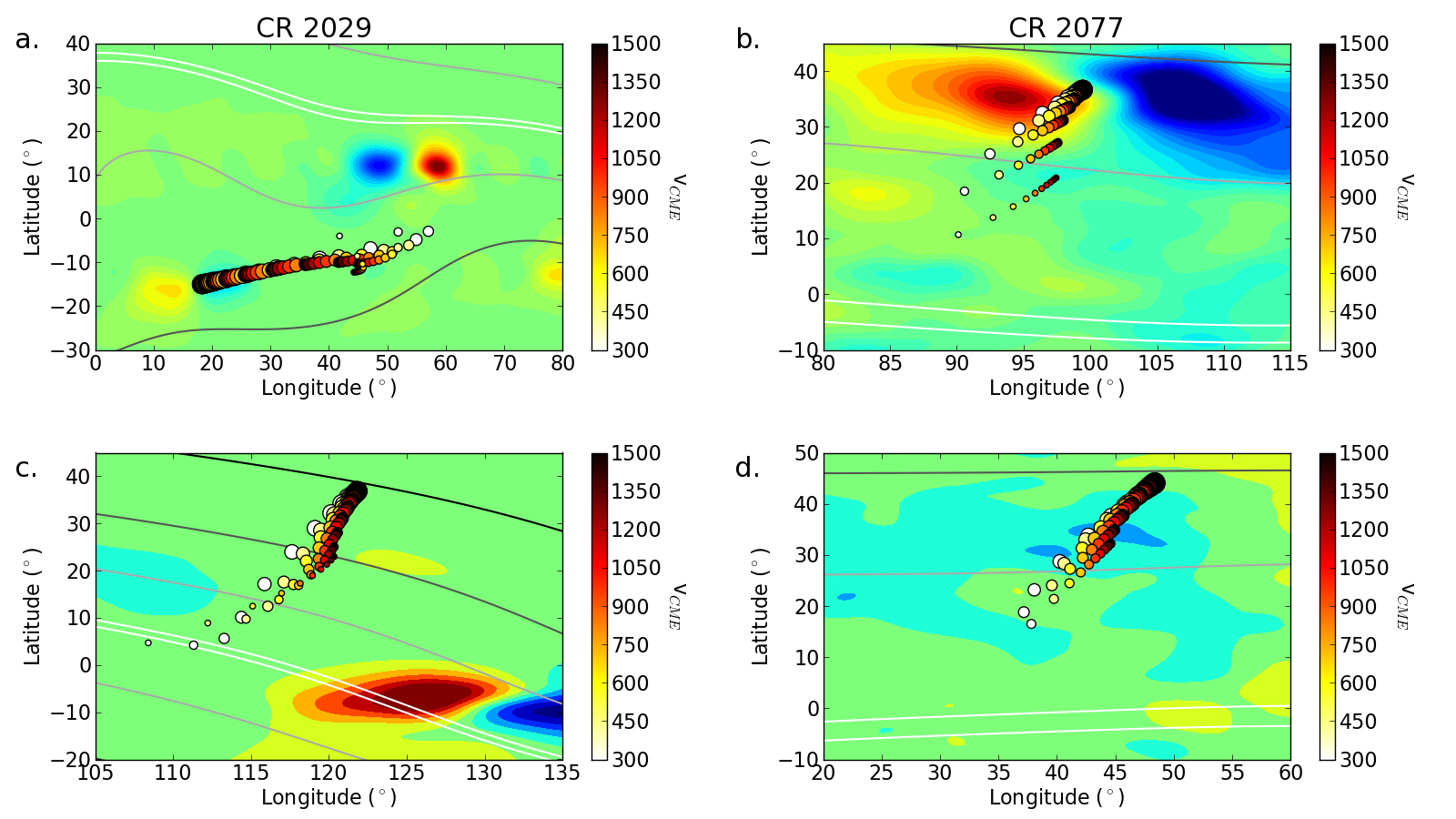}
\caption{Each panel shows the latitude and longitude at 10 $\rsun$ of 100 individual ForeCAT CMEs with varying masses and final propagation velocities.  All CMEs were initiated at the location of CME 2029A (Fig. \ref{fig:MvV}(a)), CME 2077A (Fig. \ref{fig:MvV}(b)), CME 2029b (Fig. \ref{fig:MvV}(c)), or CME 2077a (Fig. \ref{fig:MvV}(d)).  The circle size represents the CME mass (larger being more massive) and the color fill represents the CME velocity.  The background color contours show the radial magnetic field at 1.05 $\rsun$, showing the location of the ARs, and the line contours show the total magnetic field strength at the source surface height, 2.5 $\rsun$, which indicates the location of the HCS. }\label{fig:MvV}
\end{figure}

Each panel in Figure \ref{fig:MvV} shows ForeCAT results for 100 different CMEs with different masses and final propagation velocities.  The CMEs in Fig. \ref{fig:MvV}(a) and (b) were initiated at the initial locations of CME 2029A and 2077A.  All parameters other than the mass and the velocity remain unchanged from the control cases.  In Fig. \ref{fig:MvV}, the color and line contours represent the solar magnetic field at 1.05 and 2.5 $\rsun$ respectively, as in Fig. \ref{fig:CRs}.  Each circle represents the final deflected position of a CME at 10 $\rsun$, which corresponds to approximately half of the total deflection in the control cases.  We focus on the effects of the deflection forces, not the rotation of the Sun itself, so we show position in the fixed observer longitude.  The background magnetic field corresponds to their locations at the initiation of each CME.  The actual magnetic structure will have rotated to the west as the CME propagates to 10 $\rsun$.  For a 300 km s$^{-1}$ and a 1500 km s$^{-1}$ this corresponds to approximately 4$\mydeg$ and 1$\mydeg$ of rotation.  The size of the circle indicates the CME mass, with the largest circles representing the most massive CMEs.  The CME masses range between 10$^{14}$ and 10$^{16}$ g.  The color of each circle represents the final propagation speed as indicated in the color bar.

The CMEs corresponding to CME 2029A are initiated at the AR in the bottom left corner of Fig. \ref{fig:MvV}(a).  As seen for the individual cases in sections \ref{Mass} and \ref{Vel}, the most massive CMEs with the fastest radial propagation velocities deflect the least.  The fastest, most massive CME deflects 1.2$\mydeg$.  The slowest, least massive CME deflects 26.9$\mydeg$.  The largest deflection does not occur for the least massive CME because, as seen in Fig. \ref{fig:MVW}(a), the CME with mass 10$^{14}$ g exhibits a change in the longitudinal motion which brings it back toward the initial position.  For this declining phase background, the local gradients do play a role in determining the direction of the deflection, but much of the initial local imbalance is aligned with the direction of the global gradients, so that the direction of deflections do not vary significantly more than in Fig. \ref{fig:MvV}(a).  For any single CME mass, the CMEs tend to lie in a single plane and the mass and velocities affect the amount of deflection within this plane.

The CMEs in Fig. \ref{fig:MvV}(b) are initially aligned with the AR present at the top of the figure.  In general, these CMEs show the same dependence on mass and velocity but deflect less than the CR 2029 CMEs due to the weaker background magnetic field.  The deflections at 10 $\rsun$ range from 0.2$\mydeg$ for a 10$^{16}$ g, 1500 km s$^{-1}$ CME, up to 27.5$\mydeg$ for a 10$^{14}$ g, 300 km s$^{-1}$ CME.  Some variation occurs in the direction of deflection, however this is small due to the weak local gradients present during solar minimum.  The slowest CMEs show the most longitudinal deflection, which occurs as a result of the local gradients.  All CMEs move toward the magnetic minimum located at the HCS, but none have reached it by 10 $\rsun$.

Both CME 2029A and CME 2077A correspond to CMEs erupting from ARs, however, CMEs can also erupt from the quiet sun.  Figs. \ref{fig:MvV}(c) and \ref{fig:MvV}(d) present results for 100 CMEs with different mass and velocities with the initial positions of CME 2029b and 2077a, which correspond to flux ropes formed at PILs in the quiet sun.  For both CRs, the deflections show less variation in the direction of deflection than their AR counterparts.  The final positions of all the CMEs within each CR lie closer to within a single plane in three-dimensional space, the orientation of the plane determined by the direction of the global magnetic gradients.  In the quiet sun, the local magnetic gradients are significantly weaker than those near an AR so the quiet sun CMEs deflect in the direction of the global magnetic gradients.  CR 2029 has stronger magnetic field and gradients so slightly more variation is visible for that CR.  For both CRs, we see deflections of comparable magnitude to the AR CMEs despite the weaker quiet sun magnetic field.  The deflections range up to 29.2$\mydeg$ for CR 2077 and up to 34.8$\mydeg$ for CR 2029.  The quiet sun CMEs tend to originate from extended PILs and are therefore longer than the AR CMEs.  This causes a decrease in the CME density such that the weaker background magnetic field can still produce significant deflections.  For CR 2029 we see that the slowest, least massive CMEs can deflect slightly beyond the position of the HCS by 10 $\rsun$.

\section{Carrington Rotation Results}\label{CRmaps}
\begin{figure}
\includegraphics[scale=0.8]{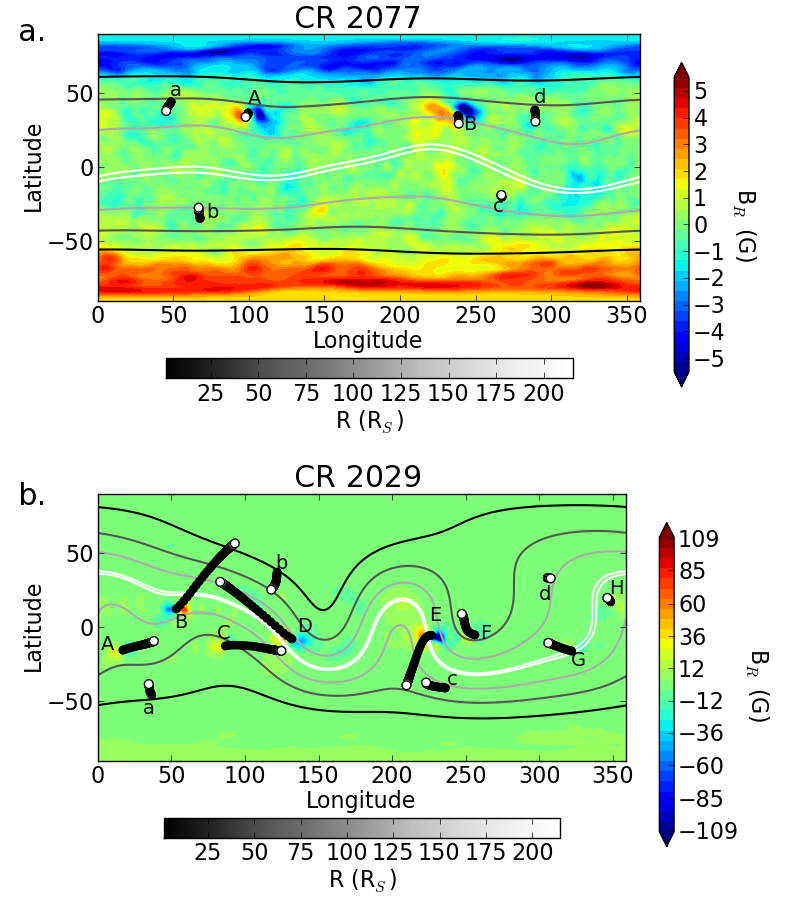}
\caption{Map showing the deflections of several CMEs for CR 2077 (\ref{fig:CRmaps}(a)) and CR 2029 (\ref{fig:CRmaps}(b)).  The color indicates a CME's radial position with black representing near the Sun and white representing the position at 1 AU.  As in Fig. \ref{fig:CRs} the color contours show the magnetic field at the solar surface, showing AR locations, and the line contours show the magnetic field at 2.5 $\rsun$, showing the HCS location.}\label{fig:CRmaps}
\end{figure}

In addition to describing variation with mass and velocity for a single location within a CR, ForeCAT shows how the deflections vary for different locations within a single CR.  Figure \ref{fig:CRmaps} shows deflections for several locations in CR 2077 (top) and 2029 (bottom).  As in previous figures, the color and line contours represent the solar magnetic field at 1.05 and 2.5 $\rsun$, respectively, at the initial locations in the fixed observer coordinate system.  Each map contains a CME deflection for each AR (upper case labels) and four additional quiet sun CMEs (lower case labels).  The initial position of each CME is listed in Table 1.  All CME's have a mass of 10$^{15}$ g and a final propagation velocity of  475 km s$^{-1}$.  Fig. \ref{fig:CRmaps} contains the full trajectory of each CME with the color indicating the radial distance: black represents near the Sun and white represents near 1 AU.  

For CR 2077, all the deflections move the CMEs closer to the magnetic minimum.  These deflections tend to be primarily latitudinal as the weaker local gradients of CR 2077 tends to cause very weak longitudinal deflections.  We find that the quiet sun CMEs can deflect as much as their counterparts during solar minimum.  In general, ARs have stronger magnetic field than the quiet sun, however, this difference is less noticeable during solar minimum.  Additionally, the quiet sun CMEs tend to be larger that the AR CMEs (column 5 in Table 1) so the decrease in density balances the slight decrease in the magnetic forces.  For both CME types, the magnitude of these deflections tends to be smaller than the declining phase deflections due to the weaker background magnetic field strength and gradients.

All but two of the CR 2029 CMEs show deflection of more than 10$\mydeg$ with the AR CME deflections typically exceeding those of the quiet sun CMEs.  Each CME initially moves towards the global minimum but the individual behaviors tend to be more complicated than seen in CR 2077 due to the increase complexity and strength of the declining phase magnetic field.  These CMEs also show significantly more longitudinal motion, which results from the inclination of the HCS and the effects of local magnetic gradients.  

CMEs 2029B, D, E, G, and H all erupt from ARs located close to the position where the HCS forms at higher distances.  These CMEs initially move toward the HCS but then exhibit a variety of behavior.  CMEs 2029B, E, and G deflect towards local null points in the streamer region and traverse the location corresponding to the HCS at heights below the height at which it actually forms.  CME 2029C deflects to the HCS and CME 2029 H does not reach the HCS by 1 AU.  CME 2029D deflects along the HCS until the distance at which the magnetic forces can no longer affect the deflection, causing the CME to continue moving in a direction tangent to the HCS.  

The quiet sun CMEs of CR 2029 show deflection of varying magnitudes.  CMEs 2029A and 2029B show deflections of at least 10$\mydeg$ toward the HCS, but neither reaches it by 1 AU.  CME 2029c also shows a significant amount of deflection, but this motion parallels the HCS rather than moving the CME toward it, indicating that local gradients can still play a significant role for quiet sun CMEs in strong global magnetic gradients.  CME 2029D shows almost no deflection.  CMEs 2029C and D have the same shape, which shows that the difference in their deflections is due to differences in the deflection forces, rather than a difference in densities. 

\section{CME Rotation}\label{CMERot}
\begin{figure}
\includegraphics[scale=0.4]{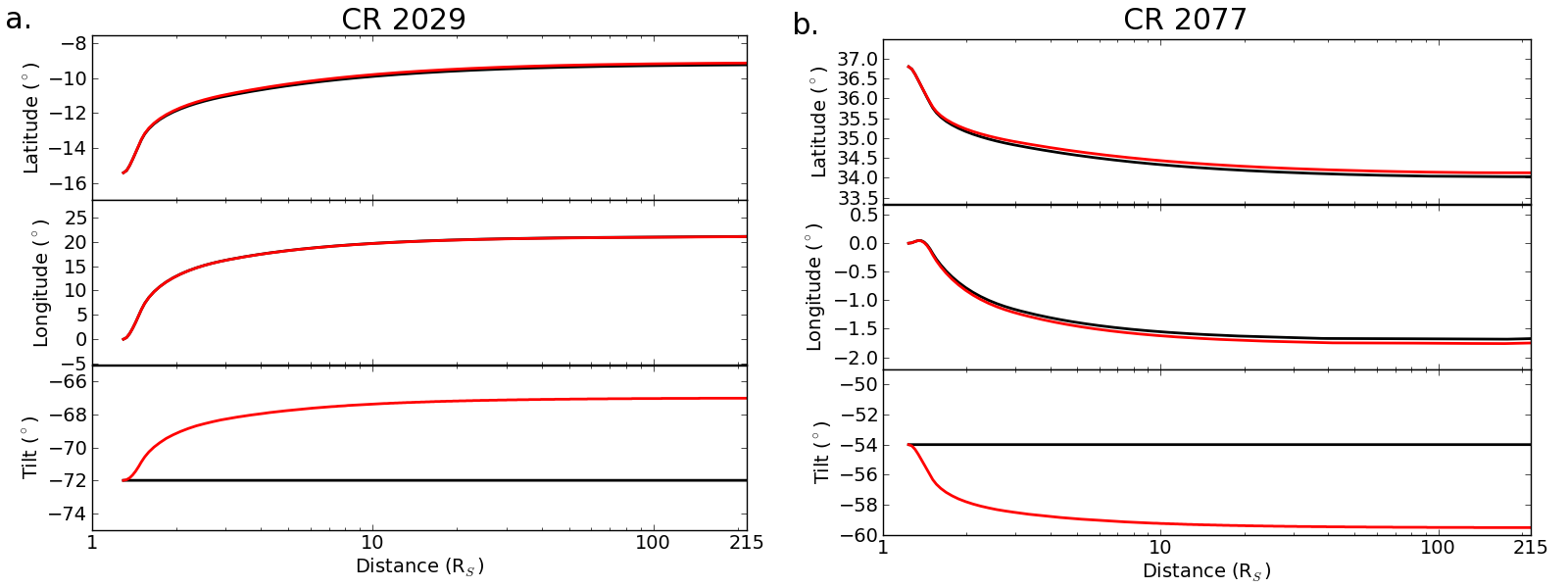}
\caption{Deflections for CMEs 2029A (Fig. \ref{fig:rot}(a)) and 2077A (Fig. \ref{fig:rot}(b)) when CME rotation is included.  The black lines show the same results as Fig. \ref{fig:CPAcontrols} and the red cases have the same initial parameters but the CME is allowed to rotate.  The bottom panels show the change in the CME tilt versus radial distance. }\label{fig:rot}
\end{figure}

ForeCAT can include the effects of CME rotation due to a torque created by differential forces along the CME's toroidal axis.  Figure \ref{fig:rot} shows results for CMEs 2029A and 2077A when the rotation is included (red lines).  The black lines represent the cases from Fig. \ref{fig:CPAcontrols} when rotation is not included.  The top two panels show the change in the latitude and longitude of the CME and the bottom panel show the CME tilt.

Both CMEs rotate of order 5$\mydeg$, however, CME 2077A rotates clockwise and CME 2029A rotates counter-clockwise.  For these cases we find that the inclusion of the rotation has little effect on the deflection - the difference between the latitude and longitude of the rotating and non-rotating cases is negligible.  It is possible that larger rotation could cause the CME to experience highly different forces from those felt in the original orientation, we show that this can occur for CME 2029D when we compare with the K13 results in Section \ref{FC1comp}.

\section{Discussion and Conclusions}\label{DisCon}
We present a new three-dimensional version of ForeCAT, which we use to study the trends in CME deflection.  We also show that torques due to differential forces along the CME can cause rotation about the CME axis.  To the best of our knowledge, ForeCAT is currently the only simple model which simulates both deflection and rotation and runs fast enough to be useful for either real time predictions or large parameter space studies.

The deflection of a CME depends on both CME parameters and the background magnetic field.  ForeCAT results for CMEs initiated in CR 2029 and CR 2077 show that the majority of the deflection occurs below 10$\rsun$.  For a strong magnetic background, such as CR 2029, the deflections are primarily determined at distances less than 2 $\rsun$.  

Our model, ForeCAT simulates the magnetic deflection using actual magnetic forces determined from the solar background but relies on simplified empirical models to describe the CME's radial propagation and expansion.  We have shown that the radial velocity model and CME size affect deflection so it is essential to develop realistic propagation and expansion models so that ForeCAT can accurately predict CME deflections.  ForeCAT's ad-hoc kinematic model can be constrained with observations and the deflection can be used to constrain CME masses, as well as the drag and background solar magnetic field \citep{Kay15}.

The magnitude and direction of the deflection is determined by CME parameters such as the mass and velocity, as well as the solar magnetic field and magnetic gradients.  Both global gradients, determined by the relative orientation of CHs and the HCS, and local gradients, related to ARs or other small scale structures, can contribute to the total deflection.  We find that wide, slow, low mass CMEs in backgrounds with strong magnetic fields and magnetic gradients (i.e. near solar maximum) tend to deflect the most and that these deflections typically have a larger longitudinal component as a result of both local and global magnetic gradients.

Many CMEs do not deflect to the magnetic minimum, the HCS on global scales.  The global magnetic gradients always cause a CME to deflect toward the magnetic minimum but the forces are not necessarily strong enough to fully deflect the CME to the magnetic minimum.  Alternatively, the forces may be sufficiently strong to push the CME beyond the HCS.  This can occur when local gradients cause the CME to deflect towards a null point in the low corona and the CME moves underneath the HCS at a height where the streamer region has not yet transitioned into the HCS.  The CME can also deflect beyond the the HCS at further distances as a result of global gradients.  When a CME reaches the HCS, the magnetic forces change direction as the CME attempts to deflect beyond it.  However, since the magnetic forces decrease rapidly with distance, if the CME reaches the HCS beyond a few solar radii the force may be insufficient to slow the CME's deflection motion.  If the CME reaches a magnetic minimum in the low corona the changing forces can halt the deflection motion and act as a potential well, trapping the CME at the magnetic minimum.

During solar minimum (CR 2077), the weak magnetic field and gradients leads to smaller deflections that tend to have very little longitudinal components.  However, during more complex backgrounds with stronger magnetic field, such as the declining phase CR 2029, the magnitude of the deflections increases and they tend to have a larger longitudinal component.  The higher inclination of the HCS causes the global magnetic gradients to have a larger longitudinal component.  Additionally, this declining phase CR has stronger local gradients that deflect the CME perpendicular to the PIL of an AR.  For CMEs that erupt from the quiet sun, the local magnetic gradients typically do not contribute significantly to the deflection so that the direction of the deflection is typically determined solely by the global gradients.

At distances where the forces are negligible, angular momentum is conserved and the deflection is proportional to the inverse of the distance.  We suggest that observed interplanetary CMEs deflections be split into two categories, those which are deflected primarily in the low corona (and thereafter having a constant angular momentum), and those which are actively accelerated at interplanetary distances (where the angular momentum is still changing). Deflections proportional to the inverse of the distance may result from a constant deflection velocity obtained in the lower corona, and not require any additional interplanetary deflection. Interplanetary deflections exceeding the inverse of the distance, however, would indicate additional deflection forces or a radial deceleration of the CME. Additionally, we expect a difference in the direction of the deflection for these two categories. CMEs deflecting mainly in the low corona can deflect in a wide range of directions based on the direction of the coronal gradients. CMEs which are actively accelerated by interplanetary forces will systematically deflect to the east or the west if the CME is faster or slower than the solar wind \citep{Wan04}.

This work has only used static magnetic fields that were derived from synoptic maps.  The global magnetic field does not vary greatly on the time scales used to map the solar magnetic field.  However, the global magnetic gradients alone do not determine the deflection, rather we see that local gradients can contribute significantly, particularly for slow, low mass CMEs in strong magnetic fields.  In addition to the inherent temporal lag associated with synoptic maps, ARs can evolve on time scales shorter than the time it takes the CME to propagate through the low corona so the local gradients should be accurately represented both spatially and temporally.

\appendix

\section{Comparison with \citet{Kay13}}\label{FC1comp}
The first version of ForeCAT determined the deflection of a CME cross-section within a fixed deflection plane, the direction of which was determined by the magnetic gradients at 2 $\rsun$.  Major changes have been incorporated to the version of ForeCAT used in this paper.  First, we abandon the deflection plane - the CME is free to deflect in 3D space.  Second, the full CME flux rope, opposed to only a cross section, is considered via a grid representing a toroidal surface.  Third, the initial CME is aligned with an polarity inversion line to determine the location and tilt.  Fourth, we use a self-similar expansion model instead of the MSOE model.  Fifth, the \citet{Che96} density model has been replaced with a modified version of the \citet{Guh06} model to better describe the three-dimensional density variations.  Sixth, we include the effects of solar rotation, including the resulting interplanetary magnetic field beyond the source surface of the PFSS model.  Finally, we can include rotation of the CME due to differential deflection forces along the toroidal axis.

\begin{figure}
\includegraphics[scale=0.8]{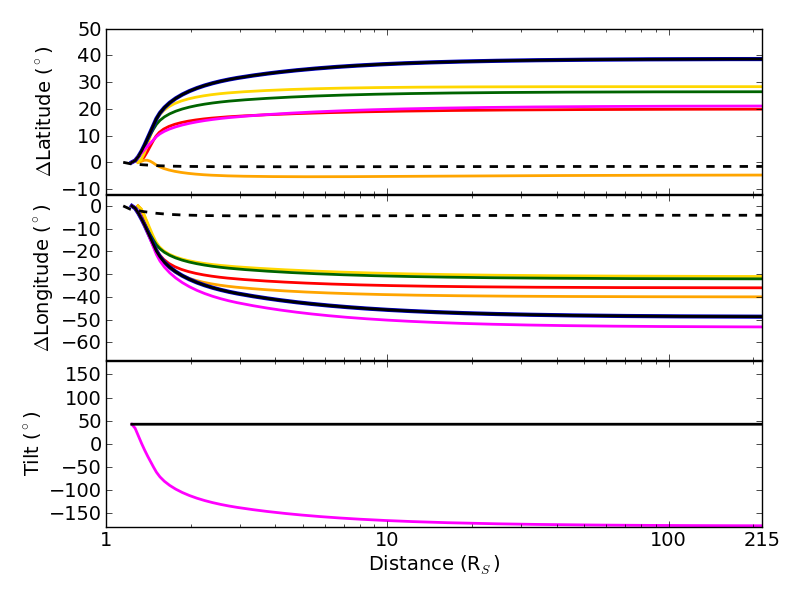}
\caption{Effect of ForeCAT model changes and comparison with the results of K13 (dashed black) line. The top panels show the change in CME latitude (top) and longitude (middle) and the bottom panel shows the change in tilt due to CME rotation.}\label{fig:FC1comp}
\end{figure}

Figure \ref{fig:FC1comp} shows the effect of each these changes on the latitudinal (top) and longitudinal (middle) deflection and comparison with the results of K13.  The case considered in K13 corresponds to CME 2029D from this work.  The dashed black line in Fig. \ref{fig:FC1comp} represents the K13 results using the PFSS model.  To replicate the K13 results as best as possible, we use a 1x3 grid (an odd number of points are required in the poloidal direction), place the CME at the same location with the same tilt as K13, use the expansion from the MSOE model and the \citet{Che96} density model, and do not include the effects of solar rotation or CME rotation.  The only differences between this version, hereafter Case 1, shown in red in Fig. \ref{fig:FC1comp}, and the K13 is the freedom from the deflection plane and the enforced decrease in the deflection velocity to conserve angular momentum.  We immediately see a significant difference in the CME deflection.  The K13 CME was forced to move in the direction of the global magnetic gradients at 2 $\rsun$, however this direction is not aligned with the direction of the local magnetic gradients lower in the corona.  The K13 CME only experienced the component of the local gradients pointing in the same direction as the global gradients.  This forced the CME to quickly move toward the magnetic minimum at the HCS, then stop its motion as its momentum carried it toward the opposite side of the magnetic potential well at the HCS.  Case 1 is free to deflect in the direction of the local magnetic gradients and begins deflecting to the northeast, nearly perpendicular to the southeastern motion of the K13 CME.  Since Case 1 experiences the full magnetic force in the low corona, opposed to the deflection plane component, it deflects more than the K13 CME.  This initial northeastern deflection causes the CME to move along the HCS at further distances, rather than moving directly into in and being trapped by the potential well.

We next consider the effect of having a higher resolution grid covering the front of the CME, rather than a single cross section near the CME nose.  Case 2, shown in orange in Fig. \ref{fig:FC1comp}, is the same as Case 1, but the grid has been changed to the standard 15x13 grid used in this work.  The change in grid causes a small increase in the longitudinal deflection but a noticeable change in the latitudinal motion.  The latitudinal motion initially begins in the same direction as before, but at a lower rate.  The latitudinal deflection then reverse and the CME moves slightly past its initial latitude.  The CME deflects according to the average force from all grid points, so increasing the number of grid points used decreases the relative contribution of any single grid point.  A large number of grid points averages out any extreme values and tends to cause smaller average forces.  The deflection tends to decrease as the number of grid point increases with the rate of change decreasing until convergence near a 15x13 grid.

CMEs in this work are initially aligned with a PIL to minimize the initial magnetic forces.  This corresponds to a slight change in the initial location from K13 and a larger change in the tilt.  Case 3 (yellow line in Fig. \ref{fig:FC1comp}) is the same as Case 2, but with the updated initial position and tilt.  This causes a minor difference in the longitudinal motion and an increase in the latitudinal deflection with the result being more similar to Case 1 than Case 2.  This change is primarily due to the 23$\mydeg$ change in the tilt of the CME.

Case 4, the green line in Fig. \ref{fig:FC1comp}, is the same as Case 3, but with self-similar expansion instead of the MSOE model.  The MSOE model causes rapid overexpansion below 2 $\rsun$ followed by a gradual decrease in the angular size of the CME at further distances.  For this case, we find very little difference between the two expansion models, with a slight decrease in the deflection for the self-similar model which expands less at small distances.

Case 5, the blue line in Fig. \ref{fig:FC1comp}, includes the modified \citet{Guh06} model instead of the \citet{Che96} density model.  This line is essentially indistinguishable from the black line representing Case 6.  We have increased the thickness of the line corresponding to Case 5 to show it lies underneath Case 6. The change in density model causes a large increase in both the latitudinal and longitudinal deflection.  The two density models agree near 1 AU, however they differ by nearly an order of magnitude in the corona.  The \citet{Guh06} model is less dense which causes a significant reduction in the total drag and leads to larger deflections.

The model matches the version used in this work when we include the effects of solar rotation (black line in Fig. \ref{fig:FC1comp}).  Inclusion of the Parker spiral causes almost no noticeable change in the deflection. By the distances as which the magnetic field has gained a significant component the magnetic forces have become too weak to strongly influence the CME's motion.

Finally, we can include CME rotation due to the differential deflection forces, which we show with the purple line in Fig. \ref{fig:FC1comp}.  The bottom panel shows the change in tilt of this CME, and comparison with the non-rotating case.  For the two cases considered in Section \ref{CMERot}, the CMEs rotated less than 10$\mydeg$ and the inclusion of rotation had little effect on the rotation.  This CME erupted from a complex environment- an AR located near the location of the HCS at further distances.  This cause the CME to rotate over 100$\mydeg$, and this causes a significant effect on the deflection.

\section{Expected Dependence of Deflection on Parameters}\label{expdep}
Using equation \ref{acceleq} we determine how the deflection should vary based on changes in either the initial CME parameters, or properties of the background solar model.  We compare these predictions with ForeCAT results in section \ref{modelsens} and \ref{CMEinputs}.

We express the CME density, $\rho_{CME}$, in terms of the CME mass, $M_{CME}$, and the CME volume determined as the product of the toroidal length, $\ell$, and the cross-sectional area, $\pi b^2$.  The non-radial CME acceleration can be approximated in terms of the non-radial displacement, $\Delta x_{nr}$ and time, $t$.
\begin{equation}
a = \frac{\Delta x_{nr}}{t^2}
\end{equation} 
We consider the amount of deflection after the CME has propagated a specified radial distance, $\Delta x_{r}$, which we relate to the time using the radial CME velocity, $v_{CME}$.  Finally, the curvature in the tension force can be approximated as $1/b$ and the magnetic gradients approximated as $B^2 / L$ where $L$ is a representative length scale of the gradients in the magnetic energy.  Solving for the non-radial displacement, we find
\begin{equation}\label{scalingeq}
\Delta x_{nr} = \frac{\pi b^2 \ell \; \Delta x_{r}^2}{M \; v_{CME}^2} \left( -\frac{B^2}{8\pi L} + \frac{B^2}{4 \pi b} - \frac{C_d \tanh \beta \; \rho_{SW}}{\pi b}v_{nr} |v_{nr}| \right)
\end{equation}
where we have assumed that the solar wind is purely radial.  This equation describes the linear deflection assuming constant forces in the time it takes the CME to propagate $\Delta x_{r}$ but we can use it to understand how the deflection varies with input parameters.  The change in the angular position of the CME depends on the linear non-radial displacement and the radial distance of the CME.

We find a straightforward dependence on the CME mass and final propagation velocity.  These terms only affect the coefficient that multiplies the force terms, rather than affecting the individual forces.  As the mass increases, the deflection decreases because the forces act upon higher densities.  As the velocity increases, the deflection decreases because the CME requires less time to travel $\Delta x_r$ and the non-radial forces act upon the CME for less time.  An increase in either the mass or velocity causes a decrease in the deflection but Eq. \ref{scalingeq} we expect the deflection to be more sensitive to velocity than mass because the displacement is proportional to $v_{CME}^{-2}$ and $M^{-1}$.

The other parameters affect the individual force terms.  We first consider the case without drag to better understand the effect on the deflection forces.  The magnetic pressure and magnetic tension terms have nearly the same form, the relative contributions depend on the ratio $2L/b$.  The length scale of the gradients will change throughout the solar cycle.  On average, this length will decrease as the solar magnetic configuration becomes more complex near solar maximum.  The length scale of the gradients also decreases with distance as the magnetic field configurations simplifies into only radial magnetic field and the HCS.  

If the cross-sectional width is significantly larger than the scale of the gradients, which will happen for wide CMEs near solar maximum or far from the Sun, then the magnetic pressure gradient term dominates the deflection.  In this case, the deflection will increase as the width increase.  The width does not affect the magnetic pressure gradient force but causes a decrease in the CME density, assuming a constant mass.  The deflection will depend strongly on the width as the displacement is proportional to $b^2$.  The deflection will also increase as the length scale of the gradients decreases since the force is proportional to $L^{-1}$.

If the cross-sectional width is significantly smaller than the scale of the gradients, which will happen for narrow CMEs close to the Sun near solar minimum, the magnetic tension term dominates the deflection force.  The net dependence of the deflection on the width is then proportional to $b$.  For an increase in the width, the decrease in the density outweighs the decrease in the tension force, causing a net increase in the deflection.

When the width and gradient length scales are comparable both the magnetic pressure gradients and magnetic tension contribute to the deflection.  The deflection will still increase as the length scale decreases, but to a lesser extent as the tension term is unaffected.  Conversely, as the width increases the tension term decreases while the magnetic pressure term remains unaffected.  The corresponding decrease in density will cause an increase in deflection that will exceed the corresponding increase in the tension-only case but be smaller than that of the gradient-only case.  For all cases, as the magnetic field increases, the deflection will increase proportional to $B^2$.  We expect larger deflections near solar maximum when the magnetic field is the strongest.  Without the effects of drag, we expect the largest deflections for wide, slow, low mass CMEs near solar maximum.

When the effects of drag are included, the amount of deflection decreases but the zeroth order scaling relations do not change.  The net force will still be smaller than the case without drag as the drag acts in the opposite direction to the velocity caused by the deflection forces.  This decrease in the amount of deflection will increase as the drag coefficient increases.  For the case where the magnetic pressure gradients exceed the tension, the deflection will still increase proportional to $b^2$ because the drag decreases proportional to $b^{-1}$ so it will have a less noticeable effect for wide CMEs.  For the tension-only case, the tension will increase as the width decreases, however the drag increases at the same rate so that the net effect will still be proportional to $b$ because of the change in density.  When both components of the deflection force contribute the deflection still increases with width, but the net deflection is smaller than in the drag free case.  Again we determine that the largest deflections should occur for wide, slow, low mass CMEs near solar minimum.

\section{Derivation of the Deflection Forces}\label{defF}
ForeCAT deflections forces result from the background solar magnetic field, $\vec{B}_{SW}$, which we assume drapes around the CME.  The precise nature of the draping of the 3D magnetic field lines onto a 2D surface is an area of open research, however, we must make some approximation to determine the deflection force.  We assume that the components of the background magnetic field in the toroidal and poloidal directions remain unchanged.  As shown in Fig. \ref{fig:diag} the toroidal direction, $\hat{t}$, parallels the long axis of the torus, and the poloidal direction, $\hat{p}$, is perpendicular to $\hat{t}$.  We assume that the component in the normal direction, $\hat{n}$, drapes in the poloidal direction as CMEs tend to be smaller in that direction relative to the toroidal direction so the background solar wind will tend to flow in the poloidal direction. 

We define the draped solar magnetic field, $B_d$, at a point on the CME surface using the toroidal and poloidal coordinate system.
\begin{equation}
\vec{B}_d = (B_p, B_t) = (\vec{B}_{SW}\cdot\hat{p} + \vec{B}_{SW}\cdot\hat{n}) \hat{p} + (\vec{B}_{SW}\cdot\hat{t}) \hat{t} 
\end{equation}
We assume that as the magnetic field drapes around the CME the ratio of the toroidal to poloidal magnetic field does not change.  This describes a geodesic path where the draped field line has no additional curvature beyond that of the CME surface.  We determine the direction of the draped magnetic field for each grid point so that the draped magnetic field is only required to follow a geodesic path local to that grid point.
 
The magnetic pressure gradient force, $F_G$ is defined as
\begin{equation}
F_{G} = \frac{\nabla_{\perp} B^2} {8 \pi}
\end{equation}
where $\nabla_{\perp}$ represents the gradient in the direction perpendicular to $\vec{B}$.  ForeCAT determines the full gradient $\nabla  B^2_{d}$ of the draped solar magnetic field by determining the magnetic field at four locations $\pm$0.5$\mydeg$ in latitude and longitude. We then remove the component parallel to $\vec{B}_{d}$.
\begin{equation}
F_{G} = \frac{\nabla B^2_{d}} {8 \pi}  - \left(\frac{\nabla B^2_{d}} {8 \pi} \cdot \hat{B}_{d} \right) \hat{B}_{d}
\end{equation}

The magnetic tension force, $F_T$ is defined as 
\begin{equation}
F_{T} = \frac{\kappa B^2} {4 \pi}
\end{equation}
which requires the curvature, $\kappa$, of the draped solar magnetic field lines, which we have assumed is equal to the curvature of the CME surface in the direction of the draping.  

We define the torus surface, $X$, in terms of the toroidal and poloidal angles, $\theta_t$ and $\theta_p$ and the shape parameters $a$, $b$, $c$, and $d$.
\begin{equation}
X (\theta_t , \theta_p) = [(a + b \cos \theta_p) \cos \theta_t, \; b\sin \theta_p, \; (c + b \cos \theta_p) \sin \theta_t]
\end{equation}
The un-normalized tangent vectors in the toroidal and poloidal directions are $X_{\theta_t}$ and $X_{\theta_p}$, where the subscripts indicate derivatives with respect to that variable.  The direction of the draped magnetic field can then be written as
\begin{equation}
\vec{v}_d = B_p X_{\theta_p} + B_t X_{\theta_t}
\end{equation}
where $B_p$ and $B_t$ are the magnitudes of the poloidal and toroidal components of the draped magnetic field.  Using the definition of the curvature of a surface we find that
\begin{equation}
\kappa = \frac{eB_p^2 + 2f B_p B_t + gB_t^2}{EB_p^2 +2FB_p B_t + GB_t^2}
\end{equation}
here $E$, $F$, and $G$ are coefficients of the first fundamental form, defined as 
\begin{eqnarray}
E & = & X_{\theta_p} \cdot X_{\theta_p} \\ 
F & = & X_{\theta_p} \cdot X_{\theta_t} \\ 
G & = & X_{\theta_t} \cdot X_{\theta_t} 
\end{eqnarray}
and $e$, $f$, and $g$ are coefficients of the second fundamental form, defined as
\begin{eqnarray}
e & = & n \cdot X_{\theta_p \theta_p} \\ 
f & = & n \cdot X_{\theta_p \theta_t} \\ 
g & = & n \cdot X_{\theta_t \theta_t} 
\end{eqnarray}
here the double subscripts indicate second derivatives.

\section{Moment of Inertia of the CME Torus}\label{MoI}
We determine an analytic form of the moment of inertia of an elliptical torus which we can evaluate from the CME shape at each time step.  We start by defining a Cartesian coordinate system centered at the $O$ in Fig. \ref{fig:diag} with the x-axis parallel to the dashed red line indicating $a$, the y-axis coming out of the page, and the z-axis parallel to the red dashed line indicating $c$.  We then convert to a cylindrical coordinate system with the polar component parallel to the xz plane ($r$ and $\theta$) and the linear axis the same as the y-axis. The moment of inertial can be rewritten in cylindrical coordinates as
\begin{equation}\label{MoIcy}
I = \int \int \int \rho \ell^2 r dy dr d \theta 
\end{equation}
but the limits of integration and L must be expressed in terms of the cylindrical coordinates.  The displacement vector, $\ell$, can be written as $\ell=\sqrt{y^2 + z^2}$, which can be rewritten in cylindrical coordinates.
\begin{equation}
L = \sqrt{y^2 + r^2 \sin^2\theta}
\end{equation}

To determine the limits of integration we require a description of the torus in the cylindrical coordinates.  The surface of the torus can be defined as the points with a distance equal to the cross-sectional radius, $b$, from the toroidal axis which runs through the center of the torus. In this cylindrical coordinate system, the surface of the torus is defined by 
\begin{equation}\label{surf}
b^2 = y^2 + (r - r_e(\theta))^2
\end{equation}
where $r_e(\theta)$ is the radius of the elliptical toroidal axis which depends on the polar angle $\theta$ as well as the semi-major axis and eccentricity, $e$.  Typically the semi-major axis is parallel to the z-axis ($c>a$), as shown in Fig. 1, however it can be parallel to the x-axis instead ($a>c$), and ForeCAT checks to ensure the appropriate value is used.  The following equation describes the more common configuration with a semi-major axis $c$.
\begin{equation}
r_e = \frac{c (1-e^2)}{1 + e \cos \theta}
\end{equation}
Rearranging equation \ref{surf} gives the maximum absolute value of $y$, $y_{max}$.  
\begin{equation}
y_{max} = \sqrt{b^2 - (r-r_e)^2}
\end{equation}
Since the integral is symmetric with respect to $y=0$ we can multiply by a factor of 2 and integrate $y$ from zero to $y_{max}$.

The remaining limits are more straightforward, we integrate $r$ between $r_e - b$ and $r_e + b$ and $\theta$ between zero and $\pi$ as we consider a half-torus.  
\begin{equation}\label{MoIfull}
I = 2 \rho \int_0^{\pi} \int_{r_e-b}^{r_e+b} \int_0^{\sqrt{b^2 - (r-r_e)^2}}  (r^2\sin^2\theta +y^2)^2 \;r \; dy \; dr \; d \theta 
\end{equation}
An outline of the integration is presented below.  Note that $r_e$ is a function of $\theta$ but we use the shorthand notation and substitute in the full expression once we get to the $\theta$ integral.
\begin{equation}\label{MoI1}
I = 2 \rho \int_0^{\pi} \int_{r_e-b}^{r_e+b} \sqrt{b^2 - (r-r_e)^2}r^3\sin^2\theta +\frac{1}{3}(b^2 - (r-r_e)^2)^{\frac{3}{2}} \;r \; dy \; dr \; d \theta 
\end{equation}
Equation \ref{MoI1} can be broken into two integrals which have the solutions
\begin{equation}
\int_{r_e-b}^{r_e+b} \sqrt{b^2 - (r-r_e)^2} \; r^3 \; d\theta = \frac{\pi}{8} r_e b^2 (4 r_e^2 + 3 b^2)
\end{equation}
and
\begin{equation}
\int_{r_e-b}^{r_e+b} (b^2 - (r-r_e)^2)^{\frac{3}{2}} \; r^3 \; d\theta = \frac{3\pi}{8} r_e b^4 
\end{equation}
which yields
\begin{equation}\label{MoI2}
I = \frac{\pi}{4} \rho b^2 \int_0^{\pi} (4r_e^3\sin^2\theta + 3 r_eb^2\sin^2\theta + r_eb^2) \; d \theta 
\end{equation}
and we must now include the full form of $r_e$.
\begin{equation}\label{MoI3}
I = \frac{\pi}{4} \rho b^2 \left[4c^3(1-e^2)^3\int_0^{\pi}\frac{\sin^2\theta\; d \theta}{(1+e \cos\theta)^3} \; + \; 3b^2c(1-e^2)\int_0^{\pi}\frac{\sin^2\theta\; d \theta}{1+e \cos\theta} \; + \; b^2c(1-e^2)\int_0^{\pi}\frac{d \theta}{1+e \cos\theta}\right]
\end{equation}
These integrals have the analytic solutions
\begin{equation}
\int_0^{\pi}\frac{\sin^2\theta\; d \theta}{(1+e \cos\theta)^3} = \frac{\pi}{2(1-e^2)^{\frac{3}{2}}}
\end{equation}
\begin{equation}
\int_0^{\pi}\frac{\sin^2\theta\; d \theta}{1+e \cos\theta} = \frac{\pi(1 - \sqrt{1-e^2})}{e^2}
\end{equation}
\begin{equation}
\int_0^{\pi}\frac{d \theta}{(1+e \cos\theta)^3} = \frac{\pi}{\sqrt{1-e^2}}
\end{equation}
which after some manipulation gives us the final analytic form of the moment of inertia.
\begin{equation}\label{MoIfinal}
I = \frac{\rho b^2 c \pi^2}{4} \left[2(1-e^2)^{\frac{3}{2}}c^2 \; + \; 3b^2(1-e^2) \frac{1-\sqrt{1-e^2}}{e^2} \; + \; b^2\sqrt{1-e^2} \right]
\end{equation}

The circular half-torus case ($e=0$) requires simpler integrals and can be derived much easier.  Taking the limit of equation \ref{MoIfinal} as $e$ tends to zero produces the same result as the circular derivation.
\begin{equation}\label{circ}
I = \frac{1}{2} \rho b^2 c \pi^2 (c^2 + \frac{5}{4}b^2)
\end{equation}


\begin{thebibliography}{}
\bibitem[Altschuler \& Newkirk(1969)]{Alt69} Altschuler, M. D., \& Newkirk, G. 1969, SoPh, 9, 131
\bibitem[Altschuler et al.(1977)]{Alt77} Altschuler, M. D., Levine, R. H., Stix, M., \& Harvey, J. 1977, SoPh, 51, 345
\bibitem[Aschwanden(2005)]{AscB} Aschwanden,  M. J. 2005, Physics of the Solar Corona. An Introduction with Problems and Solutions (2nd ed.; Berlin: Springer) 
\bibitem[Aschwanden(2009)]{Asc09} Aschwanden, M. J. 2009, Ann. Geophys. 27, 3275
\bibitem[Byrne et al.(2010)]{Byr10} Byrne, J. P., Maloney, S. A., Refojo, J. M., \& Gallagher, P. T. 2010, Nature Communications, 1, 74
\bibitem[Burlaga et al.(1981)]{Bur81} Burlaga, L., Sittler, E., Mariani, F., \& Schwenn R. 1981, JGR, 86, A8, 6673
\bibitem[Cane \& Richardson(2003)]{CR03} Cane, H. V., \& Richarson, I. V. 2003, JGR, 108, A4, 1156
\bibitem[Cargill et al.(1996)]{Car96} Cargill, P. J., Chen, J., Spicer, D. S., \& Zalesak, S. T. 1996, JGR, 101, A3
\bibitem[Cargill(2004)]{Car04} Cargill, P. J. 2004, SoPh, 221, 135
\bibitem[Chen(1996)]{Che96} Chen, J. 1996, JGR, 101, A12, 27519
\bibitem[Chen et al.(1997)]{Che97} Chen, J., Howard, R. A., Brueckner, G. E., et al. 1997, ApJ, 490, L191
\bibitem[Chen et al.(2000)]{Che00} Chen, J., Santoro, R. A., Krall, J., et al. 2000, ApJ, 533, 1, 481
\bibitem[Cremades \& Bothmer(2004)]{Cre04} Cremades, H., \& Bothmer, V. 2004, A\&A, 422, 307
\bibitem[Evans et al.(2008)]{Eva08} Evans, R. M., Opher, M., Manchester, W. B., \& Gombosi, T. I. 2008, ApJ, 687, 1355
\bibitem[Evans et. al(2012)]{Eva12} Evans, R. M., Opher, M., Oran, R., et al. 2012, ApJ, 756, 155
\bibitem[Fisher \& Munro(1984)]{Fis84} Fisher, R. R., \& Munro, R. H. 1984, ApJ, 280, 428
\bibitem[Forbes et al.(2006)]{For06} Forbes, T. G., Linker, J. A., Chen, J., et al. 2006, SSRv, 123, 251
\bibitem[Gibson \& Fan(2006)]{Gib06} Gibson, S. E., \& Fan, Y. 2006, ApJ, 637, L65
\bibitem[Gibson \& Low(1998)]{Gib98} Gibson, S. E., \& Low, B. C. 1998, ApJ, 493, 460
\bibitem[Gopalswamy et al.(2009)]{Gop09} Gopalswamy, N., Yashiro, S., Michalek, G., et al. 2009, EM\&P, 104, 295
\bibitem[Guhathakurta et al.(1996)]{Guh96} Guhathakurta, M., Holzer, T. E., \& MacQueen, R. M. 1996, ApJ, 458, 817
\bibitem[Guhathakurta et al.(2006)]{Guh06} Guhathakurta, M., Sittler, E. C., \& Ofman, L. 2006, JGR, 111, A11215
\bibitem[Gui et al.(2011)]{Gui11} Gui, B., Shen, C., Wang, Y., et al. 2011, SoPh, 271, 11
\bibitem[Iju et al.(2013)]{Iju13} Iju, T., Tokumaru, M., \& Fujiki, K. 2013, SoPh, 288, 1, 331
\bibitem[Isavnin et al.(2013)]{Isa13} Isavnin, A., Vourlidas, A., \& Kilpua, E. K. J. 2013, SoPh, 284, 203
\bibitem[Kay et al.(2013)]{Kay13} Kay, C., Opher, M., \& Evans, R. M. 2013, ApJ, 775, 5
\bibitem[Kay et al.(2015)]{Kay15} Kay. C, dos Santos, L. F. G., Opher, M. 2015, ApJL, accepted
\bibitem[Kilpua et al.(2009)]{Kil09} Kilpua, E. K. J., Pomoell, J., Vourlidas, A., et al. 2009, Ann. Geophys., 27, 4491
\bibitem[Klein \& Burlaga(1982)]{KB82} Klein, L. W., \& Burlaga, L. F. 1982, JGR, 87, 613
\bibitem[Liu et al.(2010a)]{Liu10a} Liu, Y., Davies, J. A., Luhmann, J. G., et al. 2010a, ApJL., 710, 82
\bibitem[Liu et al.(2010b)]{Liu10b} Liu, Y., Thernisien, A., Luhmann, J. G., et al. 2010b, ApJ, 722, 1762
\bibitem[Liu et al.(2011)]{Liu11} Liu, Y. C.-M., Opher, M., Wang, Y., \& Gombosi, T. I. 2011, A\&A, 527, A46
\bibitem[Lugaz et al.(2011)]{Lug11} Lugaz, N., Downs, C., Shibata, K., et al. 2011, 738, 127
\bibitem[Lugaz et al.(2012)]{Lug12} Lugaz, N., Farrugia, C. J., Davies, J. A., et al. 2012, 759, 68
\bibitem[Lynch \& Edmondson(2013)]{Lyn13} Lynch, B. J. \& Edmondson, J. K. 2013, ApJ, 764, 1, 87
\bibitem[Low(1982)]{Low82} Low, B. C. 1982, ApJ, 254, 796
\bibitem[Low(1984)]{Low84} Low, B. C. 1984, ApJ, 281, 392
\bibitem[Lugaz et al.(2010)]{Lug10} Lugaz, N., Hernandez-Charpak, J. N., Roussev, I. I., et al. 2010, ApJ, 715, 493
\bibitem[MacQueen et al.(1986)]{Mac86} MacQueen, R. M., Hundhausen, A. J., \& Conover, C. W. 1986, JGR, 91, 31
\bibitem[Maloney \& Gallagher(2010)]{Mal10} Maloney, S. A. \& Gallagher, P. T 2010, ApJL, 724, 2, L127
\bibitem[Mann et al.(2003)]{Man03} Mann, G., Klassen, A., Aurass, H., Classen, H. T. 2003, A\& A., 400, 329
\bibitem[Mierla et al.(2011)]{Mie11} Mierla, M., Inhester, B., Rodriguez, L., et al. 2011, JASTP, 73, 1166
\bibitem[M{\"o}stl et al.(2012)]{Mos12} M{\"o}stl, C., Farrugia, C. J., Kilpua, E. K. J., et al. 2012, ApJ, 758, 10
\bibitem[Nieves-Chinchilla et al.(2012)]{NC12} Nieves-Chinchilla, T., Colaninno, R., Vourlidas, A., et al. 2012, JGR, 117, A06106
\bibitem[Nieves-Chinchilla et al.(2014)]{NC14} Nieves-Chinchilla, T., Vourlidas, A., Stenborg, G., et al. 2014, ApJ, 779, 55
\bibitem[Panasenco et al.(2011)]{Pan11} Panasenco, O., Martin, S., Joshi, A. D., \& Srivastava, N. 2011, JASTP, 73, 1129
\bibitem[Panasenco et al.(2013)]{Pan13} Panasenco, O., Martin, S. F., Velli, M., \& Vourlidas, A. 2013, SoPh, 287, 391
\bibitem[Patsourakos et al.(2010a)]{Pat10a} Patsourakos, S., Vourlidas, A., \& Stenborg, G. 2010, ApJL., 724, L188
\bibitem[Patsourakos et al.(2010b)]{Pat10b} Patsourakos, S., Vourlidas, A., \& Kliem, B. 2010, A\&A 522, A100
\bibitem[Richardson \& Cane(2010)]{Ric10} Richardson, I. G., \& Cane, H. V. 2010, SoPh, 264, 189
\bibitem[Riley et al.(2006a)]{Ril06a} Riley, P., Linker, J. A., Miki\'c, Z., et al. 2006, ApJ, 653, 1510
\bibitem[Riley et al.(2006b)]{Ril06b} Riley, P., Schatzman, C., Cane, H. V., Richardson, I. G., \& Gopalswamy, N. 2006 ApJ, 647, 648
\bibitem[Schatten et al.(1969)]{Sch69} Shatten, K. H., Wilcox, J. M., Ness, N. F. 1969, SoPh, 6, 442
\bibitem[Siscoe et al.(2006)]{Sis06} Siscoe, G. L., Crooker, N. U., \& Elliot, H. A. 2006, SoPh, 239, 293
\bibitem[Shen et al.(2011)]{She11} Shen, C., Wang, Y., Gui, B., Ye, P., \& Wang, S. 2011, SoPh, 269, 389
\bibitem[Shen et al.(2012)]{She12} Shen, Y., Liu, Y., \& Su, J. 2012, ApJ, 750, 12
\bibitem[Temmer et al.(2011)]{Tem11} Temmer, M., Rollet, T., M{\"o}stl, C, et al. 2011, ApJ, 743, 2, 101
\bibitem[Thernisien et al.(2006)]{The06} Thernisien, A. F. R., Howard, R. A., \& Vourlidas, A. 2006, ApJ, 652, 763
\bibitem[Titov \& D\'{e}moulin(1999)]{TD99} Titov, V. S. \& D\'{e}moulin, P. 1999, A\&A, 351, 707
\bibitem[T\'oth et al.(2012)]{Tot12} T\'oth, G., van der Holst, B., Sokolov, I. V., et al. 2012, J. Comput. Phys., 231, 870
\bibitem[Tripathi et al.(2009)]{Tri09} Tripathi, D., Gibson, S. E., Qui, J., et al. 2009, A\&A, 498, 1, 295
\bibitem[Tripathi et al.(2012)]{Tri12} Tripathi, D., Reeves, K. K., Gibson, S. E., Srivastava, A., \& Joshi, N. C. 2012, ApJ, 778, 2, 142
\bibitem[Vandas et al.(2002)]{Van02} Vandas, M., Odstricil, D., \& Watari, S. 2002, JGR, 107, A9, 1236
\bibitem[van der Holst et al.(2010)]{vdH10} van der Holst, B., Manchester, W. B., IV, Franzin, R. A., et al. 2010, ApJ, 725, 1373
\bibitem[Wang et al.(2004)]{Wan04} Wang, Y., Shen, C., Ye., P., \& Wang, S 2004, SoPh, 222, 329
\bibitem[Wang et al.(2006)]{Wan06} Wang, Y., Xue, X., Shen. C., et al. 2006, ApJ, 646, 625
\bibitem[Wang et al.(2014)]{Wan14} Wang, Y., Wang, B., Shen, C., Shen, F., \& Lugaz, N. 2014, JGR Space Phys., 119, 5117
\bibitem[Wood et al.(2009)]{Woo09} Wood, B. E., Howard, R. A., Thernisien, A., Plunkett, S. P., \& Socker, D. G. 2009, SoPh, 259, 163
\bibitem[Xie et al.(2004)]{Xie04} Xie, H., Ofman, L., \& Lawrence, G. 2004, JGR, 109, A03109
\bibitem[Xue et al.(2005)]{Xue05} Xue, X. H., Wang, C. B., \& Dou, X. K. 2005, JGR, 110, A08103
\bibitem[Yashiro et al.(2004)]{Yas04} Yashiro, S., Gopalswamy, N., Michalek, G., et al. 2004, JGR, 109, A07105
\bibitem[Yashiro et al.(2008)]{Yas08} Yashiro, S., Michalek, G., \& Gopalswamy, N. 2008, AnGeo, 26, 31037
\bibitem[Zhang \& Dere(2006)]{Zha06} Zhang, J., \&  Dere, K. P. 2006, ApJ, 649, 1100
\bibitem[Zhou \& Feng(2013)]{Zho13} Zhou, Y. F. \& Feng, X. S. 2013, JGR Space Phys., 118, 10, 6007
\bibitem[Zhou et al.(2014)]{Zho14} Zhou, Y. F., Feng, X., \& Zhao, X. 2014, JGR Space Phys., accepted
\bibitem[Zuccarello et al.(2012)]{Zuc12} Zuccarello, F. P., Bemporad, A., Jacobs, C., et al. 2012, ApJ, 744, 1, 66
\end{thebibliography}
\end{document}